\documentclass[twocolumn]{aastex63}

\usepackage{natbib}
\usepackage{graphicx}

\newcommand{\code}[1]{\texttt{#1}}

\shorttitle{Turning AGN bubbles into radio relics with mergers}
\shortauthors{ZuHone et al.}

\begin{document}

\title{How merger-driven gas motions in galaxy clusters can turn AGN
  bubbles into radio relics}

\correspondingauthor{John ZuHone}
\email{john.zuhone@cfa.harvard.edu}

\author[0000-0003-3175-2347]{John A. ZuHone}
\affiliation{Center for Astrophysics $\vert$~Harvard~and~Smithsonian \\
Cambridge, MA 02138, USA}

\author{Maxim Markevitch}
\affiliation{NASA/Goddard Space Flight Center \\
Greenbelt, MD 20771, USA}

\author{Rainer Weinberger}
\affiliation{Center for Astrophysics $\vert$~Harvard~and~Smithsonian \\
Cambridge, MA 02138, USA}

\author{Paul Nulsen}
\affiliation{Center for Astrophysics $\vert$~Harvard~and~Smithsonian \\
Cambridge, MA 02138, USA}
\affiliation{ICRAR, University of Western Australia, 35 Stirling Hwy., 
Crawley, WA 6009, Australia}

\author{Kristian Ehlert}
\affiliation{Leibniz-Institut f\"ur Astrophysik Potsdam (AIP), An der
Sternwarte 16, 14482 Potsdam, Germany}

\begin{abstract}
Radio relics in galaxy clusters are extended synchrotron sources
produced by cosmic-ray electrons in the $\mu$G magnetic field. Many
relics are found in the cluster periphery and have a cluster-centric,
narrow arc-like shape, which suggests that the electrons are
accelerated or re-accelerated by merger shock fronts propagating
outward in the intracluster plasma. In the X-ray, some relics do
exhibit such shocks at the location of the relic, but many do not. We
explore the possibility that radio relics trace not the shock fronts
but the shape of the underlying distribution of seed relativistic
electrons, lit up by a recent shock passage. We use
magnetohydrodynamic simulations of cluster mergers and include bubbles
of relativistic electrons injected by jets from the central AGN or from 
an off-center radio galaxy. We show that the merger-driven gas motions (a) 
can advect the bubble cosmic rays to very large radii, and (b) spread the 
relativistic seed electrons preferentially in tangential direction -- along 
the gravitational equipotential surfaces -- producing extended,
filamentary or sheet-like regions of intracluster plasma enriched with
aged cosmic rays, which resemble radio relics. Once a shock front passes
across such a region, the sharp radio emission edges would trace the
sharp boundaries of these enriched regions rather than the front. We
also show that these elongated cosmic ray features are naturally
associated with magnetic fields stretched tangentially along their
long axis, which could help explain the high polarization of relics.
\end{abstract}


\section{Introduction}\label{sec:intro}

During a merger of galaxy clusters, the kinetic energy of the
colliding plasma clouds (the intracluster medium, ICM) is dissipated
into heat via shock waves and turbulence, with a small fraction of it
diverted into amplification of magnetic fields and acceleration of
cosmic ray (CR) particles that coexist within thermal
plasma. Synchrotron radio emission from the ICM indicates the presense
of ultrarelativistic cosmic ray electrons with $\gamma \sim 10^3-10^4$
and magnetic fields of $B \sim 0.1-10~\mu$G.  The brightest sources
are the radio lobes associated with active galactic nuclei (AGN),
particularly at the centers of massive clusters. Jets from the central
black hole (BH) fill these lobes, inflating giant cavities as they
displace the hot ICM, which emits in the X-ray band.

Other diffuse radio sources have no related compact source and their
nature is less clear. These sources fall into broad categories of
radio halos and radio relics \citep{vw19}. Radio halos are
volume-filling, unpolarized, and diffuse, and are thought to arise
from the reacceleration of CRe with $\gamma \sim 10^2$ (which have
long cooling times) by cluster turbulence driven by mergers
\citep{bru07,brunetti14,zuh13}. ``Giant'' halos are seen in clusters
undergoing major mergers, with typical sizes of $\sim$1~Mpc, while
``mini-halos'' are seen in the cool cores of relaxed clusters.  Radio
relics are peripheral, elongated (often in the shape of
cluster-centric narrow arcs), strongly polarized sources. Some are
associated with shock fronts in the ICM seen in X-rays. Radio relics
are further subdivided into ``gischts'', ``phoenices'', and ``AGN
relics'' \citep{kemp04,vw19}. AGN relics are fossil plasma from
AGN-blown lobes that is passively cooling. Phoenices are presumed to
arise from reacceleration of the aged CRe that originally filled the
radio lobes and were subsequently mixed into the ICM by bulk motions
and turbulence; a few sources that fit this scenario have been
discovered recently at low radio frequencies with {\em LOFAR}\/
\citep{vw19}.

Radio gischts in \citet{kemp04} terminology, or ``radio
shocks'' in \citet{vw19} terminology, are often observed
at large distances from the cluster center ($\sim1-2$~Mpc). Many are
very long ($\sim$Mpc), narrow arcs concentric with the cluster, such
as the Sausage relic \citep{vanweeren10,hoang17}, and some have
counterparts on the opposite side of the cluster.  It has been assumed
that they mark the location of shock fronts (perhaps from a cluster
merger) propagating in the ICM, which should accelerate the ICM
electrons to ultrarelativistic energies via first-order Fermi
process. The high-$\gamma$ electrons should cool rapidly after the
shock passage, resulting in narrow ($\sim$100 kpc) radio
features. This simple scenario is problematic on energy grounds --- if
the observed power spectrum of the relativistic electrons is
extrapolated down to thermal energies, it would imply an energy in
relativistic electrons comparable to the plasma thermal energy
\citep[e.g.,][]{macario11} and thus an implausibly high acceleration
efficiency for the low-Mach (${\cal M} \lesssim 3$) cluster merger
shocks. The ICM should contain a population of long-lived relativistic
CRe with lower $\gamma \sim 10^2$ from past acceleration events
\citep{sarazin99}; it is more likely that shocks re-accelerate those
electrons \citep{markevitch05,kang16}.

However, there are other problems with radio relics occurring at the
location of the shocks. While a number of relics do exhibit ICM shocks
in the X-ray exactly at the relic ``front edges'' as expected in the
simple acceleration (or re-acceleration) picture
\citep[e.g.,][]{giacintucci08,finoguenov10,macario11,shimwell15}, others
do not. For example, at the location of the most prominent and
well-studied CIZA\,J2242.8+5301 Sausage relic \citep{vanweeren10},
there is no shock in either {\em Chandra}\/ or {\em XMM}\/ X-ray images
\citep{ogrean13,ogrean14,markevitch20}. A shock with the 
required $M\approx 2.7-4.6$ \citep{hoang17} would have
been seen in those high-resolution images as easily as that in the
Bullet cluster ($M\sim 3$). It is also difficult to explain the morphology of the
Sausage relic. A shock front propagating outward is roughly a
spherical shell; if it passes though a field of uniformly distributed
seed electrons and re-accelerates them, the radio relic would look
like an umbrella in projection rather than a sausage
\citep{vanweeren11}. Furthermore, the Sausage relic is highly
polarized (50--60\% polarization fraction, \cite{vanweeren10}). Assuming the
magnetic field in the ICM is tangled at random, plasma compression by
a shock passage would not be able to create such a highly ordered
post-shock field. Indeed, a $M\sim3$ shock front in the Bullet cluster
produces enhanced radio emission but no detectable post-shock
polarization \citep{shimwell14}. Some of these issues may be potentially explained 
by the presence of MHD turbulence upstream of the shock, as noted by \citet{dom20}. 

We wish to explore an alternative possibility that radio relics such
as the Sausage trace the nonuniform distribution of the seed electrons
rather than the ICM shocks. We imagine that the ICM at the
cluster periphery contains regions polluted with aged relativistic
electrons, that those regions have sharp boundaries and shapes as long
arched filaments or sheets concentric with the cluster, and that the
magnetic fields within those regions are stretched along the filaments. 
A shock passage across such a region would create a radio relic 
with all the observed relic properties. We may catch the shock front while
it crosses such a region, creating a radio relic at the shock position. 
When the shock moves out (but not too far, in order to satisfy the 
electron cooling timescale constraint), the relic 
``front edge'' would delineate the sharp boundary of the polluted region. 
This is essentially a ``phoenix'' scenario in the \citet{vw19} terminology,
except more ordered, bigger, brighter and found further out in the clusters.
Unlike the shock-compressed relativistic fossils proposed by
\citet{ensslin01}, these regions would be dominated by thermal plasma
with the relativistic electrons providing a small fraction of the
total pressure (though higher than elsewhere in the cluster), so shock
propagation would not be affected by the relativistic component.

We know that the aged relativistic electrons should be distributed
non-uniformly in the ICM and that their distribution affects the radio
features. For example, two X-ray shocks in the Bullet cluster have 
similar Mach numbers but produce very different radio features --- one
produces a bright radio relic \citep{shimwell15}, while the other a
much fainter edge of the radio halo \citep{shimwell14}. The difference
is the radio galaxy near the relic that is likely to have enriched the
local ICM with seed electrons and possibly even connected to the relic.
Other examples where a radio galaxy apparently feeds the relic have 
been observed \citep[e.g.,][]{vanweeren17}.

Here, we want to see if gas motions in a merging cluster could
naturally produce such narrow, relic-shaped regions enhanced with relativistic
particles. To test this possibility, we perform magnetohydrodynamical
simulations of galaxy cluster mergers with bubbles, which would be produced
either by the central AGN or a radio galaxy at a larger radius.
We will show that in some cases, the gas motions can be fast and widespread
enough to advect the bubble material from the cluster core to very large radii,
and produce large, elongated ($\sim$1~Mpc), cluster-centric CR-enriched regions
that resemble radio relics such as the Sausage. We also show that in these
regions, magnetic fields are naturally stretched tangentially along
the filaments, helping to explain the large degree of polarization in relics.

We perform this study using two methods. In the first, we fire jets from a black
hole at the center of the cluster, which produces highly magnetized buoyant 
bubbles near the cluster center which then rise. In the second method, we evacuate 
a single bubble outside the cluster core (but still much closer to the cluster 
center than where the peripheral relics are found), to represent the possibility 
that a local radio galaxy injects CRe into the ICM at that location. In both cases, the 
bubbles begin to rise but are also advected by the gas motions, and eventually their
material is mixed with the ICM. Using a passive scalar field which represents the CRe, 
we then follow the evolution of this material over time to determine which features
are produced. 

This paper is organized as follows. Section \ref{sec:methods}
describes the details of the setup of the simulations, including the
cluster mergers, jets, and bubbles. In Section \ref{sec:results}, we
present our results. In Section \ref{sec:summary} we summarize our
results and present our conclusions. We assume a $\Lambda$CDM
cosmology with $h = 0.7$, $\Omega_m = 0.3$, and $\Omega_\Lambda = 0.7$.
    
\section{Methods}\label{sec:methods}

We present two types of binary cluster merger simulations in this work: one with
a central AGN in the main cluster which is modeled by a bi-directional jet, and
another within which a single bubble is evacuated at a distance from the
cluster center. The setup of the initial profiles and orbit of the two clusters,
which is the same for both simulation types, is discussed first, followed by the
details particular to each simulation. 

\subsection{Initial Condition Setup}\label{sec:ICs}

In all simulations, the gas in each cluster is modeled as a magnetized, fully
ionized ideal fluid with $\gamma = 5/3$ and mean molecular weight $\mu = 0.6$,
which is in hydrostatic equilibrium with a virialized dark matter halo which
dominates the mass of the cluster. Both clusters are then situated within a larger
computational domain at a distance from each other and given initial velocities.
We use the same method for generating these initial conditions for these mergers
as in previous works \citep{AM06,zuh10,zuh16,zuh18,zuh19}. We outline this
method in brief here. 
    
Our merging clusters consist of a large, ``main'' cluster, and a smaller infalling
subcluster. For the dark matter density profile of the clusters we have chosen a
\citet{her90} profile:
\begin{equation}\label{eqn:hernquist}
\rho_{\rm DM}(r) = \frac{M_0}{2{\pi}a^3}\frac{1}{(r/a)(1+r/a)^3}
\end{equation}
where $M_0$ and $a$ are the scale mass and length of the DM halo. The Hernquist
profile has the same dependence on radius in the center as the well-known NFW
profile \citep{NFW97}, $\rho_{\rm DM} \propto r^{-1}$ as $r \rightarrow 0$, but
is used here instead because it is more analytically tractable and its mass
profile converges as $r \rightarrow \infty$. 

For the gas density, we use a phenomenological formula which can model cool-core
clusters with temperature decreasing towards the cluster center \citep{AM06}:
\begin{equation}
\rho_{\rm gas}(r) = \rho_{g0}\left(1+\frac{r}{a_c}\right)\left(1+\frac{r/a_c}{c}\right)^\alpha\left(1+\frac{r}{a}\right)^\beta,
\end{equation}
with exponents
\begin{equation}
\alpha \equiv -1-n\frac{c-1}{c-a/a_c},~\beta \equiv 1-n\frac{1-a/a_c}{c-a/a_c},
\end{equation}
where $0 < c < 1$ is a free parameter that characterizes the depth of the
temperature drop in the cluster center and $a_c$ is the characteristic radius of
that drop, or the ``cooling radius''. We set $n = 5$ in order to have a constant
baryon fraction at large radii, and we compute the value of $\rho_0$ from the
constraint $M_{\rm gas}/M_{\rm DM} = \Omega_{\rm gas}/\Omega_{\rm DM} = 0.12$.
With this density profile and Equation \ref{eqn:hernquist}, the corresponding
gas temperature can be derived by imposing hydrostatic equilibrium.
                       
We employ two different merger scenarios in our study, which vary only in the
properties of the subcluster. In the first simulation (called ``Merger1''), the
mass ratio of the two clusters is $R \equiv M_1/M_2 = 5$ and the subcluster only
possesses dark matter. This setup produces relatively gentle sloshing gas
motions that persist for many Gyr. The second simulation (called ``Merger2'')
has $R = 3$ and the subcluster is also a cool-core cluster. This simulation
produces very fast and turbulent gas motions.         
    
The mass of the main cluster is $M_1 = 1.25 \times 10^{15}~M_\odot$ in both
simulations. To scale the initial profiles for the two subclusters, the
combination $M_i/a_i^3$ in Equation \ref{eqn:hernquist} is held constant. For
the main cluster, we chose $a_1$ = 600~kpc, $c_1$ = 0.17, and $a_{c1}$ = 60~kpc,
to resemble mass, gas density, and temperature profiles typically observed in
massive, relaxed cool-core clusters. In the ``Merger2'' simulation, its gas
properties are set assuming $c_1 = c_2$ and $a_{c1}/a_1 = a_{c2}/a_2$. We will
describe the post-merger properties of the two simulations in detail in Section
\ref{sec:merger_gas}.
            
For all simulations, the clusters start at a separation of $d$ = 3~Mpc,
and with an initial impact parameter $b$ = 500~kpc. The initial cluster
velocities are chosen so that the total kinetic energy of the system is set to
half of its potential energy, under the approximation that the objects are point
masses. This results in a relative velocity of $v_{\rm rel} \sim
1100~$km~s$^{-1}$. The initial cluster centers and velocity vectors are
situated within the $x-y$ plane of the simualation, and the mergers proceed 
entirely within the $x-y$ plane.
       
Finally, the gas in each cluster is also magnetized. To set up the initial
magnetic field we follow the procedure from \citet{brz19} and references
therein. Briefly, we set up a divergence-free turbulent magnetic field on a
uniform grid with a Kolmogorov spectrum which is isotropic in the three spatial
directions. The average magnetic field strength is scaled to be proportional to
the square root of the thermal pressure such that $\beta = p_{\rm th}/p_B$ is
constant. The magnetic field components are then interpolated from this grid onto
the cells in each simulation. 

\subsection{Simulation Code Details}\label{sec:code}

We used two different codes to carry out the simulations in this work,
mainly for historical reasons--we began with simple bubble simulations using the 
\code{FLASH} code, but later determined that a more robust physical model of 
bubble production by AGN jets would be useful. Since the \code{AREPO} code has
such a model, it was chosen to do the rest of the simulations.

\subsubsection{\code{AREPO} Jet Simulations}\label{sec:arepo_jets}

The \code{AREPO} code \citep[][]{springel10} employs a finite-volume Godunov
method on an unstructured moving Voronoi mesh to evolve the equations of
magnetohydrodynamics (MHD), and a Tree-PM solver to compute the self-gravity
from gas and dark matter. The magnetic fields are evolved on the moving mesh
using the Powell 8-wave scheme with divergence cleaning employed in
\citep{pakmor13} and in the IllustrisTNG simulations \citep{mar18}. These
simulations also include dark matter particles, which make up the bulk of the
cluster's mass and only interact with each other and the gas via gravity. 

The particle and Voronoi cell properties in the simulations are set by the
initial profiles described in \ref{sec:ICs}, with their initial velocities set
to zero in the rest frame of their cluster. The initial magnetic field is set
according to the procedure outlined above in Section \ref{sec:ICs} assuming
$\beta = 100$. The DM particles all have the same mass. The gas cells are
initialized to all have the same mass, though they are allowed to undergo mesh
refinement and derefinement during the simulation evolution, so this condition
will not remain strictly true in their case as the simulation progresses. For
each of the particle/cell positions, a random deviate $u = M(<r)/M_{\rm total}$
is uniformly sampled in the range [0,1]. Given $u$, inverting the function
$M(<r)/M_{\rm total}$ gives the radius of the particle/cell from the center of
the halo. 

For the DM particles, their initial velocities are determined using the
procedure outlined in \citet{kaz04}, which computes the particle speed
distribution function directly using the method of \citet{edd16}. Once the
particle radii and speeds are determined, positions and velocities are
determined by choosing random unit vectors in $\Re^3$.

The \code{AREPO} simulations are set within a cubical computational domain
of width $L = 40$~Mpc on a side, though for all practical purposes the region of
interest is confined to the inner $\sim 10$~Mpc. Before each simulation, we also
perform a mesh relaxation step for $\sim$100 timesteps to prevent spurious gas
density and pressure fluctuations from forming due to the random nature of the
initial condition. 

The gas cells in each simulation initially have mass $m_{\rm gas} = 1.14 \times
10^7 M_\odot$, though these can change with refinements and derefinements of
cells. The DM particles have mass $m_{\rm DM} = 9.95 \times 10^7 M_\odot$. The
main cluster in both simulations has $10^7$ DM particles, and initially has
$10^7$ gas cells. The subcluster in the ``Merger1'' simulation contains $1.97
\times 10^6$ DM particles. The subcluster in the ``Merger2'' simulation
contains $3.57 \times 10^6$ DM particles, and initially has $3.61 \times 10^6$
gas cells. The gravitational softening length for the gas cells and the DM
particles is 2~kpc.     

For simulating the effects of AGN in the \code{AREPO} simulations we use the method of
\citet{weinberger2017}. This method injects a bi-directional jet which is
kinetically dominated, low density, and collimated. Kinetic, thermal, and
magnetic energy is injected into two small spherical regions a few kpc from the
location of a black hole particle. The material injected by the jet is marked by 
a passive tracer field $\rho_{\rm CR}$ (which represents the CR-enriched material) and is advected
along with the fluid for the rest of the simulation. For the purposes of this preliminary 
study, other effects which are important to the evolution of CRe, such as diffusion, 
streaming, (re)acceleration, and cooling are ignored, but will be pursued in later
works. 
    
A black hole particle is placed at the cluster potential minimum. This particle
has a mass of $M_{\rm BH} = 3 \times 10^{9}~M_\odot$. Since we perform no
accretion onto the black hole particle in our simulations, the precise value of
the mass is of little importance, with the exception that it must be large
enough to not be affected significantly by interactions with DM particles and
remain at the cluster potential minimum.

The black hole particle serves as the site of the jet injection by the AGN. In
each simulation, the jets are fired with a power $P_{\rm jet} = 3.169 \times
10^{45}$~erg~s$^{-1}$ for a duration of $t_{\rm jet} = 100$~Myr, so the
resulting total energy injected is $E_{\rm jet} = 10^{61}$~erg in each
direction, which is a sum of kinetic, thermal, and magnetic energy. In the jet
region, the magnetic and thermal pressures are equal ($\beta_{\rm jet} = P_{\rm th}/P_B =
1$), and the injected magnetic field is purely toroidal.

\subsubsection{\code{FLASH} Bubble Simulations}\label{sec:flash_bubbles}

The \code{FLASH} code \citep{fry00,dub09,fry10} evolves the MHD equations on a 
structured adaptive mesh refinement (AMR) grid, a method of partitioning a grid 
throughout the simulation box such that higher resolutions (smaller cell sizes) 
are only used where needed. On this mesh, \code{FLASH} uses a directionally 
unsplit staggered mesh algorithm \citep[USM;][]{lee09}. The USM algorithm used 
in \code{FLASH} is based on a finite-volume, high-order Godunov scheme combined 
with a constrained  transport method (CT), which guarantees that the evolved 
magnetic field satisfies the divergence-free condition \citep{eva88}. In our 
simulation,  the order of the USM algorithm corresponds to the Piecewise-Parabolic 
Method (PPM) of \citet{col84}. The gas properties are set by the initial profiles on
the AMR cells described in Section \ref{sec:ICs}. In this simulation, the initial
magnetic field is set according to the procedure outlined above in Section
\ref{sec:ICs} assuming $\beta = 200$. 
  
The dark matter halos, and the gravitational potentials they produce, are
modeled by two rigid \citet{her90} potentials which move under the influence of
each other's gravity. The gravtiational acceleration from the halos is evaluated
by finite-differencing the potential on the AMR grid. Because the main cluster
is held fixed at the center of the simulation domain in this simulation, an
additional acceleration term is added to the grid to account for non-inertial
frame effects. This procedure is outlined in more detail in \citet{zuh11} and
\citet{rod12}. The \code{FLASH} simulation is set within a cubical computational
domain of width $L = 8$~Mpc on a side, with a finest cell size of $\Delta{x} =
0.98$~kpc.

\begin{figure*}
\centering
\includegraphics[width=0.47\textwidth]{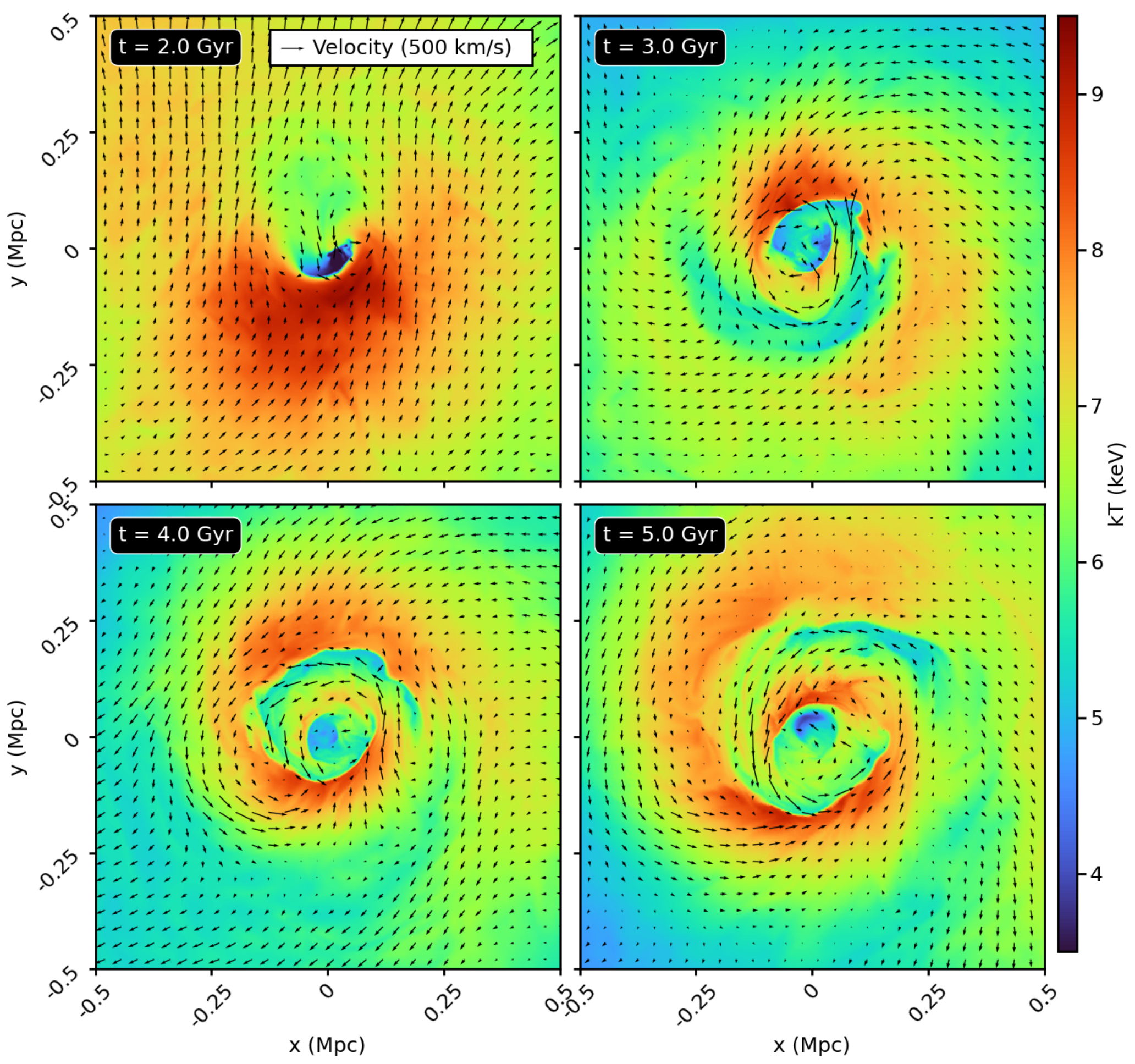}
\includegraphics[width=0.48\textwidth]{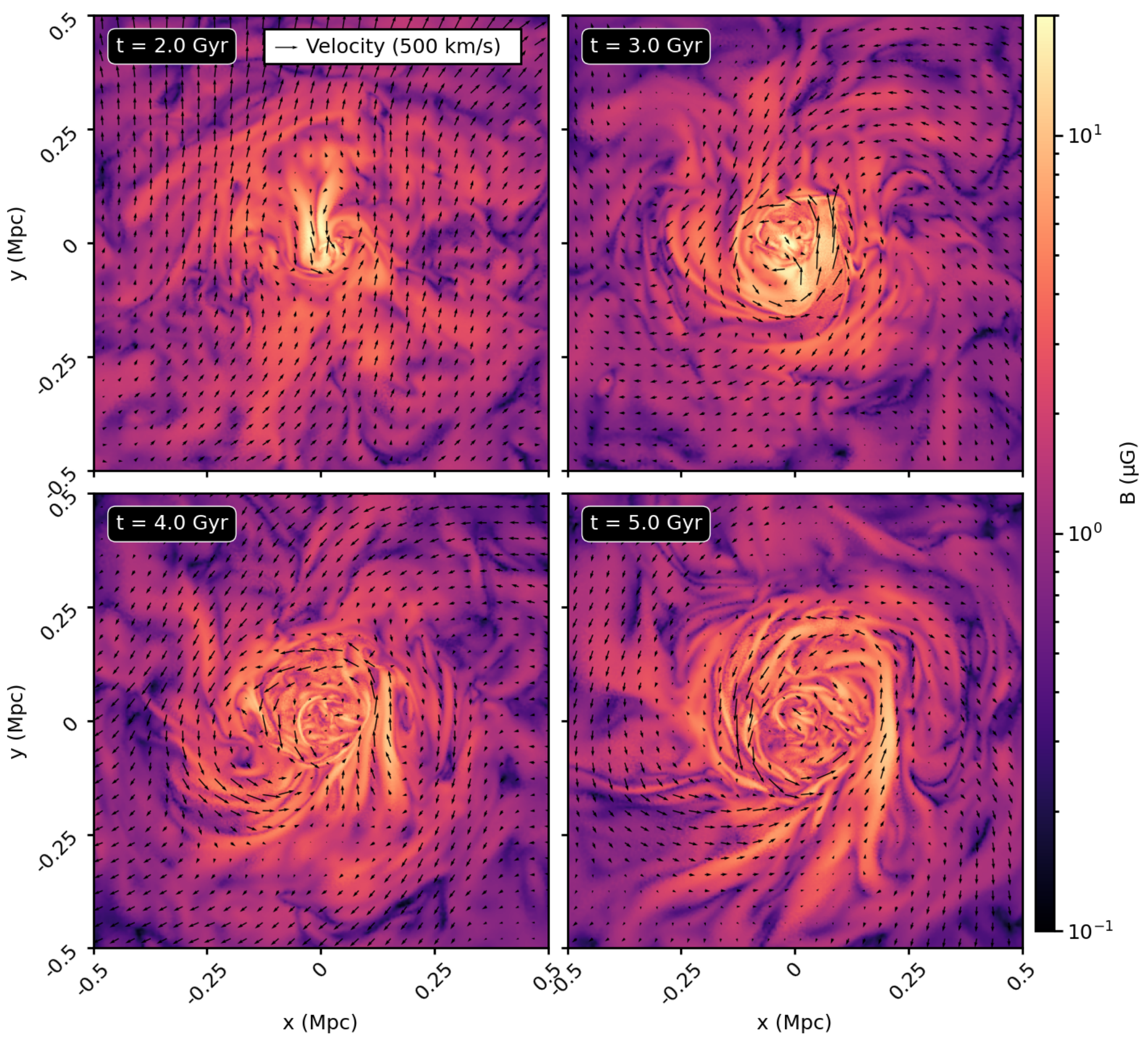}
\caption{Slices of the gas temperature (left) and the magnetic field strength
(right) for four different epochs in the ``Merger1'' simulation, centered on
the cluster potential minimum. Velocity vectors in the slice plane are also
shown. The AGN is turned on at the last snapshot shown here.
\label{fig:R5_b500_slice}}
\end{figure*}

\begin{figure*}
\centering
\includegraphics[width=0.47\textwidth]{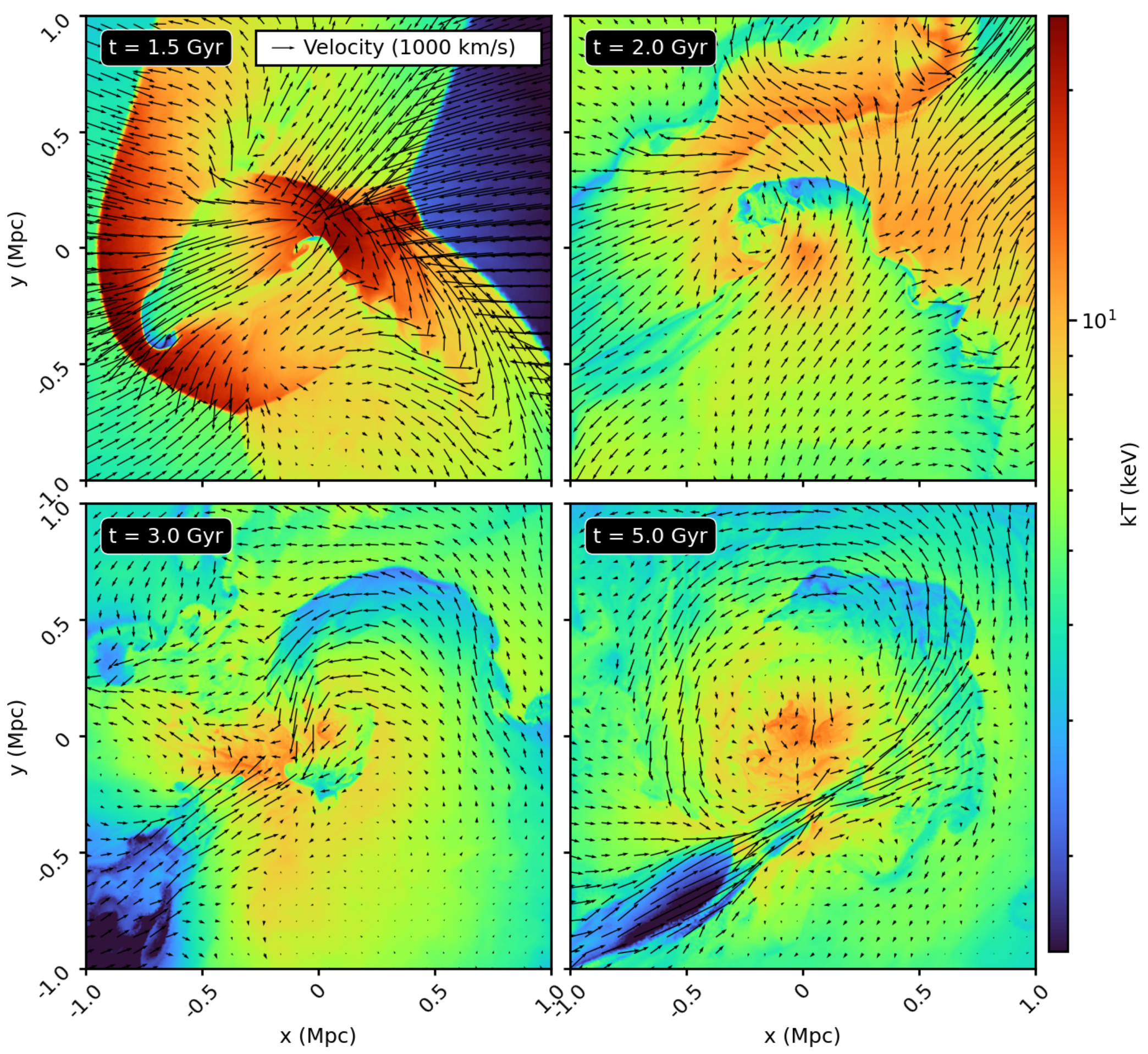}
\includegraphics[width=0.48\textwidth]{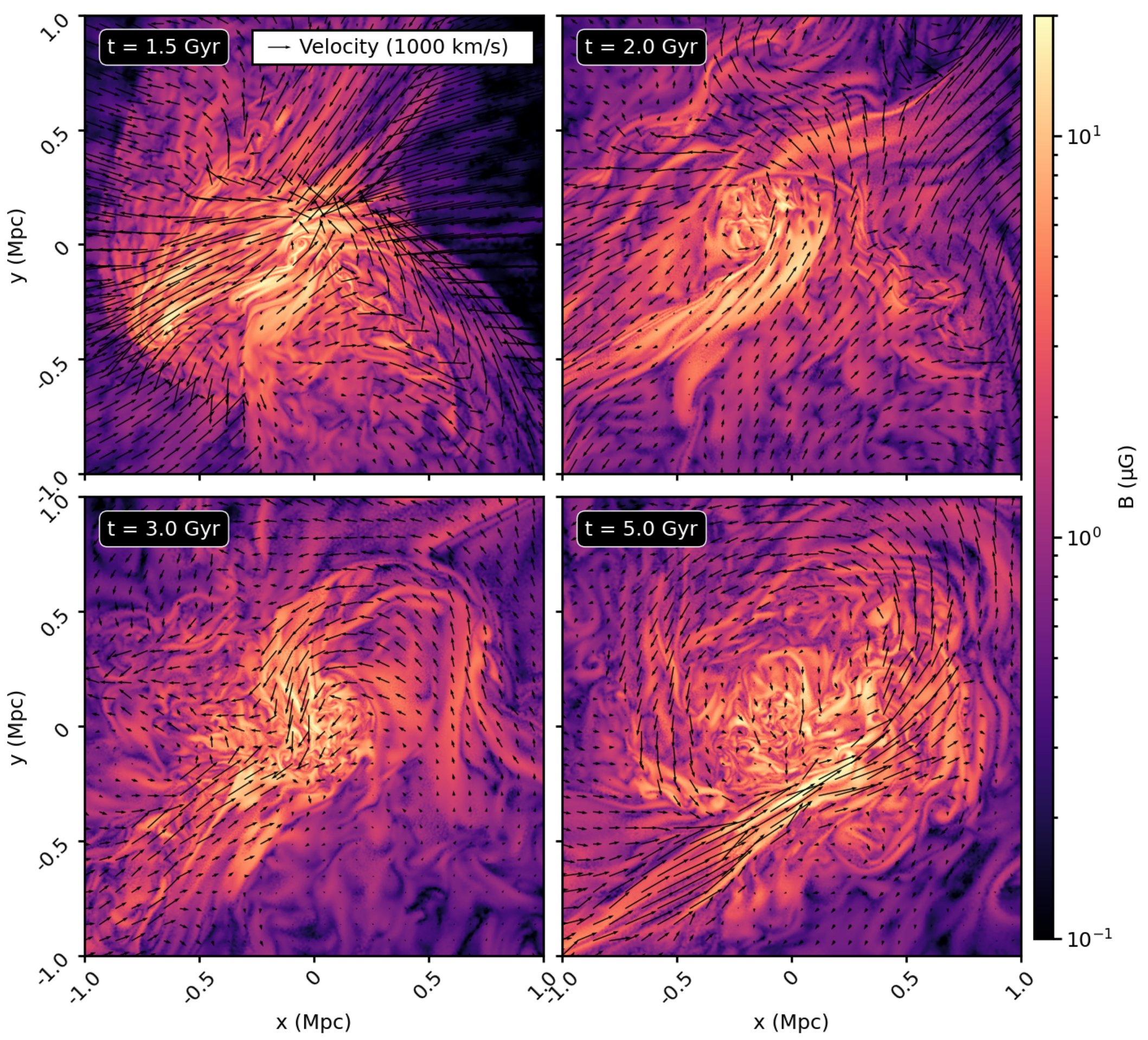}
\caption{Slices of the gas temperature (left) and the magnetic field strength
(right) for four different epochs in the ``Merger2'' simulation, centered on
the cluster potential minimum. Velocity vectors in the slice plane are also
shown.\label{fig:R3_b500_slice}}
\end{figure*}
  
In the \code{FLASH} simulation, a bubble of radius $r_{\rm bub} = 25$~kpc 
is created at a distance of $d_{\rm bub} = 200$~kpc from the cluster center.
This may represent a bubble created by an AGN at the cluster center which is
extremely stable and rises to this height, or more likely a high-entropy cloud
of hot plasma and CR injected by another radio galaxy in the cluster at that
distance. The bubble is produced by reducing the density within $r_{\rm bub}$
and adjusting the temperature such that the pressure within the bubble cells
remains constant. This results in a bubble energy of $E_{\rm bub} \sim 2 \times
10^{59}$~erg. The resulting entropy of the material in the bubble is 
$K_{\rm bub} = k_BT_{\rm bub}n_{e,{\rm bub}}^{-2/3} = 4000$~keV~cm$^2$, higher 
than the surrounding ICM by a factor of $\sim$4, so that it will buoyantly rise. 
The magnetic field in the bubble is not changed. Similar to the \code{AREPO} 
simulations, the bubble material is also marked with a passive tracer 
$\rho_{\rm CR}$ (which represents the CR-enriched material) to follow its 
subsequent evolution by its being advected along by the gas motions. Our past 
experience with simulations of a CR bubble advected by plasma in a sloshing 
cluster core \citep{zuh13} suggests that the exact initial shape of the CR cloud, 
a spherical bubble in this case, is forgotten on a couple of dynamic timescales
and has little effect the final morphology.
     
\section{Results}\label{sec:results}
        
\subsection{Characteristics of the Merger-Driven Cluster Gas Properties}\label{sec:merger_gas}

Before describing the simulations with the effects of AGN bubbles included, we
first detail the properties of the gas within each cluster as a result of the
merging activity itself. 
    
The ``Merger1'' simulation is the same setup as presented in a number of
previous works \citep{AM06,zuh10,zuh16,zuh18,zuh19}. A gasless, DM-only
subcluster perturbs the center of a relaxed, cool-core cluster and produces
sloshing motions. Slices through the merger plane of the temperature and the
magnetic field strength for four epochs of this simulation after the first core
passage of the subcluster are shown in Figure \ref{fig:R5_b500_slice}. The
sloshing motions develop cold fronts and gas velocities of several hundred
km~s$^{-1}$ in the core region. By the time the jets are switched on at $t$ =
5~Gyr, the largest cold front has propagated out to $r \sim 250$~kpc, and most
of the region inside $r \lesssim 500$~kpc is caught up in the subsonic spiral
motions. These motions have also amplified the magnetic field within the
sloshing region \citep[as shown originally in][]{zuh11} and wound the field
lines in a mostly tangential direction. We run this simulation for both the
\code{AREPO}/jet and \code{FLASH}/bubble cases, though we only show the merger
evolution of the former case in Figure \ref{fig:R5_b500_slice}.

\begin{figure*}
\centering
\includegraphics[width=0.98\textwidth]{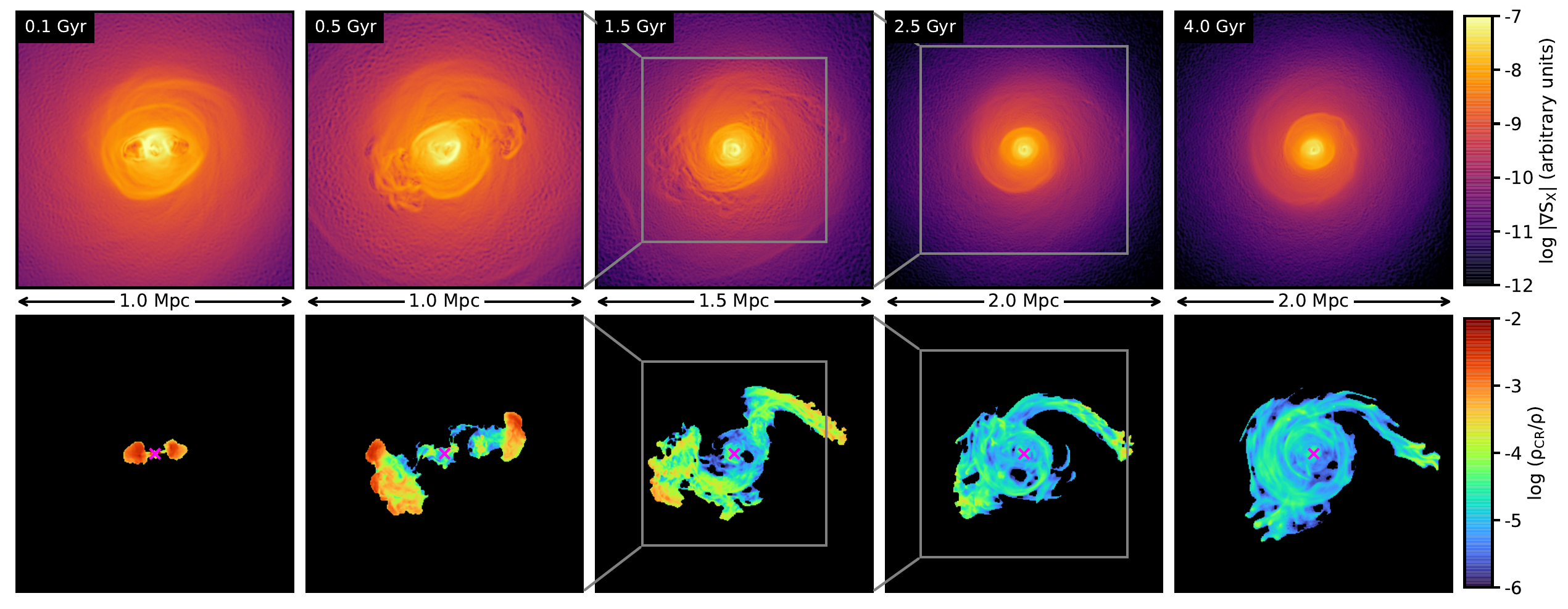}
\caption{The evolution of the ``Merger1''/jet simulation for the first 4~Gyr
after jet ignition, with jets aligned along the $x$-axis. The top panels show
projected X-ray surface brightness gradient, and the bottom panels show projected CR
fraction. The physical width of each panel expands from left to right, indicated
by the scale bars and the inset boxes. The magenta ``x'' marks the cluster
center in the bottom panels. All projections are along the
$z$-axis.\label{fig:R5_b500_jets_x}}
\end{figure*}
  
\begin{figure*}
\centering
\includegraphics[width=0.98\textwidth]{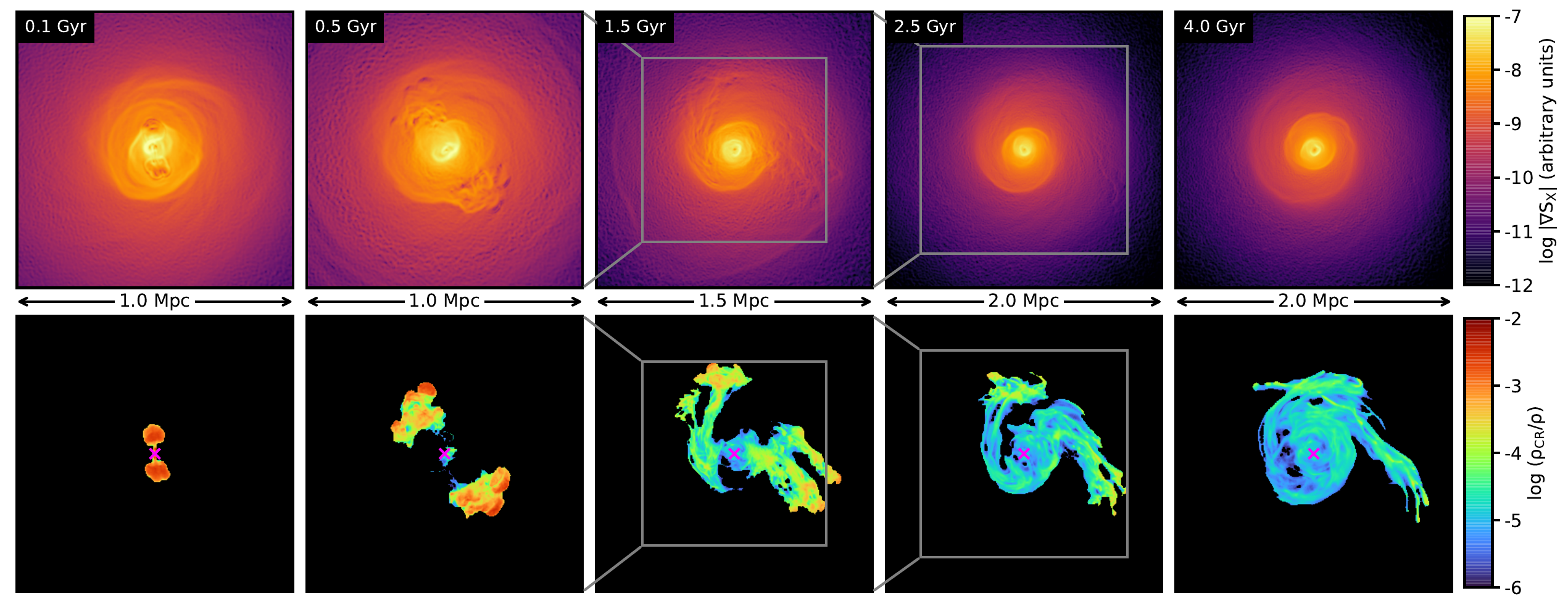}
\caption{The evolution of the ``Merger1''/jet simulation for the first 4~Gyr
after jet ignition, with jets aligned along the $y$-axis. The time labels show $\tau
= t - t_{\rm AGN}$, where $t_{\rm AGN}$ = 5~Gyr. The top panels show
projected X-ray surface brightness gradient, and the bottom panels show projected CR
fraction. The physical width of each panel expands from left to right, indicated
by the scale bars and the inset boxes. The magenta ``x'' marks the cluster
center in the bottom panels. All projections are along the $z$-axis.  
\label{fig:R5_b500_jets_y}}
\end{figure*}

\begin{figure*}
\centering
\includegraphics[width=0.98\textwidth]{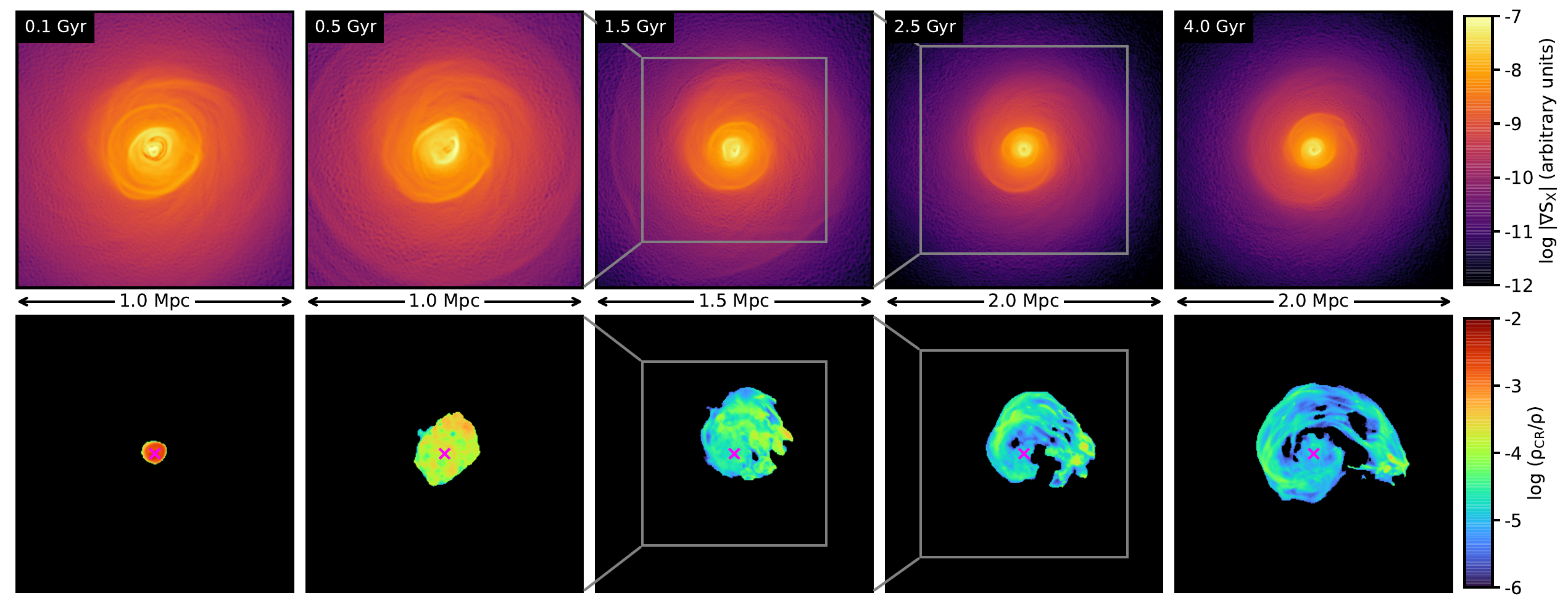}
\caption{The evolution of the ``Merger1''/jet simulation for the first 4~Gyr
after jet ignition, with jets aligned along the $z$-axis. The time labels show $\tau
= t - t_{\rm AGN}$, where $t_{\rm AGN}$ = 5~Gyr. The top panels show
projected X-ray surface brightness gradient, and the bottom panels show projected CR
fraction. The physical width of each panel expands from left to right, indicated
by the scale bars and the inset boxes. The magenta ``x'' marks the cluster
center in the bottom panels. All projections are along the $z$-axis.  
\label{fig:R5_b500_jets_z}}
\end{figure*}
  
\begin{figure*}
\begin{center}
\includegraphics[width=0.98\textwidth]{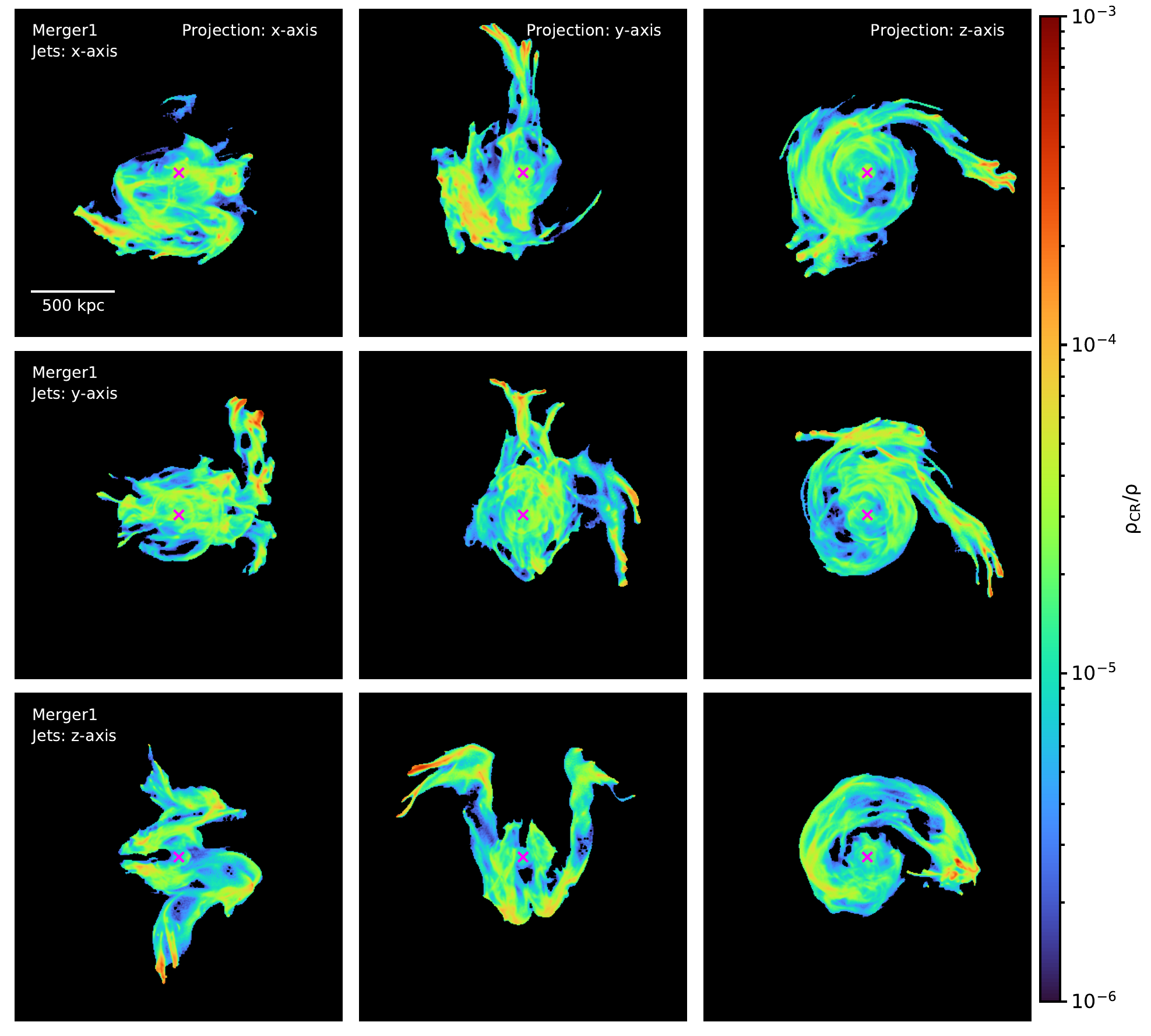}
\caption{The projected CR distribution along the principle axes of the
simulation box for the ``Merger1''/jet simulations, for all cases of the jet
direction, at $\tau$ = 4~Gyr. The position of the cluster center is marked by a
magenta ``x''.\label{fig:R5_b500_cr_proj}}
\end{center}
\end{figure*}
  
Slices through the merger plane of the temperature and the magnetic field
strength for four epochs of the ``Merger2'' simulation after the first core
passage of the subcluster are shown in Figure \ref{fig:R3_b500_slice}. In this
simulation, a gas-filled subcluster only three times smaller than the main
cluster passes near the cluster core, driving shocks and large-scale bulk motions
up to $\sim$1000~km~s$^{-1}$ in the core region. These motions produce large
cold fronts which propagate to large radii of $r \sim 700$~kpc. At later times
($t \sim 3-5$~Gyr), the subcluster makes its second core passage, driving a
fast, highly magnetized, cold inflow of gas to the south of the core. The gas
motions and magnetic field in the core region are highly turbulent by $t =
5.0$~Gyr.     
    
\subsection{Evolution of the Spatial Distribution of the Bubble Material}

We implement the effects of AGN in all of the merger simulations at $t = 5$~Gyr,
$\Delta{t} \sim 3.5$~Gyr after the first core passage. For the description of
the simulations from this point on, the times will be measured in terms of $\tau
= t - t_{\rm AGN}$ where $t_{\rm AGN}$ = 5~Gyr. At this point, in all
simulations the gas motions are well-developed, as shown in the previous section. 
In the jet simulations, we explore three different orientations of the jet axis, 
along the three principal axes of the simulation box, $x$, $y$, and $z$. As a 
reminder, the merger occurs in the $x$-$y$ plane.

\subsubsection{The ``Merger1''/jet Simulations}

The evolution of the jet-injected material in the ``Merger1'' simulations is
shown in Figures \ref{fig:R5_b500_jets_x} (jets in the $x$-direction),
\ref{fig:R5_b500_jets_y} (jets in the $y$-direction), and
\ref{fig:R5_b500_jets_z} (jets in the $z$-direction). The top panels in these 
figures (and following ones) show the gradient of X-ray surface brightness as 
computed using the Gaussian Gradient Magnitude (GGM) technique \citep{sanders2016,walker2016}. 
This allows sharp features such as cold fronts and bubbles to be shown more clearly. 
At the end of the jet injection at $\tau$ = 0.1~Gyr, two bubbles with diameters 
of $\sim$70~kpc, separated by $\sim$130~kpc, are visible as cavities in the X-ray and as
enhancements in the projected bubble distribution (left-most panels of Figures
\ref{fig:R5_b500_jets_x}-\ref{fig:R5_b500_jets_z}). By $\tau$ =
0.5~Gyr the bubbles have risen to radii of $\sim$200-300~kpc, and have dragged cold, 
dense, and highly magnetized gas from the core outward in their wakes. The sloshing 
motions have already noticeably displaced the bubbles away from the original jet axis, 
seen most obviously in the cases of jets being launched within the merger plane
(Figures \ref{fig:R5_b500_jets_x} and \ref{fig:R5_b500_jets_x}).
At this point, the bubbles are starting to become disrupted by gas turbulence and 
instabilities, and their material (and the gas they have entrained behind them) is 
beginning to mix with the surrounding ICM. As the simulation proceeds to 
$\tau$ = 1.5~Gyr, the sloshing gas motions begin to spread the bubble material 
around in a tangential direction.

At later stages ($\tau \gtrsim 1.5$~Gyr in Figures \ref{fig:R5_b500_jets_x}-
\ref{fig:R5_b500_jets_z}), the bubble material's rise is significantly slowed as 
it has become thoroughly mixed with the cluster ICM and its entropy is lowered. 
The merger-driven gas motions now dominate its evolution, resulting in two effects. 
The first is that the central volume of $\sim$200-300~kpc bounded by the sloshing cold
fronts is filled uniformly with bubble material. This effect is interesting, as
it provides another reason why we may expect radio mini-halo emission to be
confined by cold fronts if the CRe are re-accelerated by turbulence, as first
noted by \citet{maz08} and simulated in \citet{zuh13}. Secondly, the outermost
gas motions produce long, relatively straight or arc-like strips of bubble material which
extend and evolve to larger radii of $\sim$500-700~kpc. These strips have sharp
boundaries reminiscent of radio relics. In the plane of the sky parallel to the
gas motions, these strips have a width of $\sim$100-200~kpc. 
    
\begin{figure*}
\centering
\includegraphics[width=0.98\textwidth]{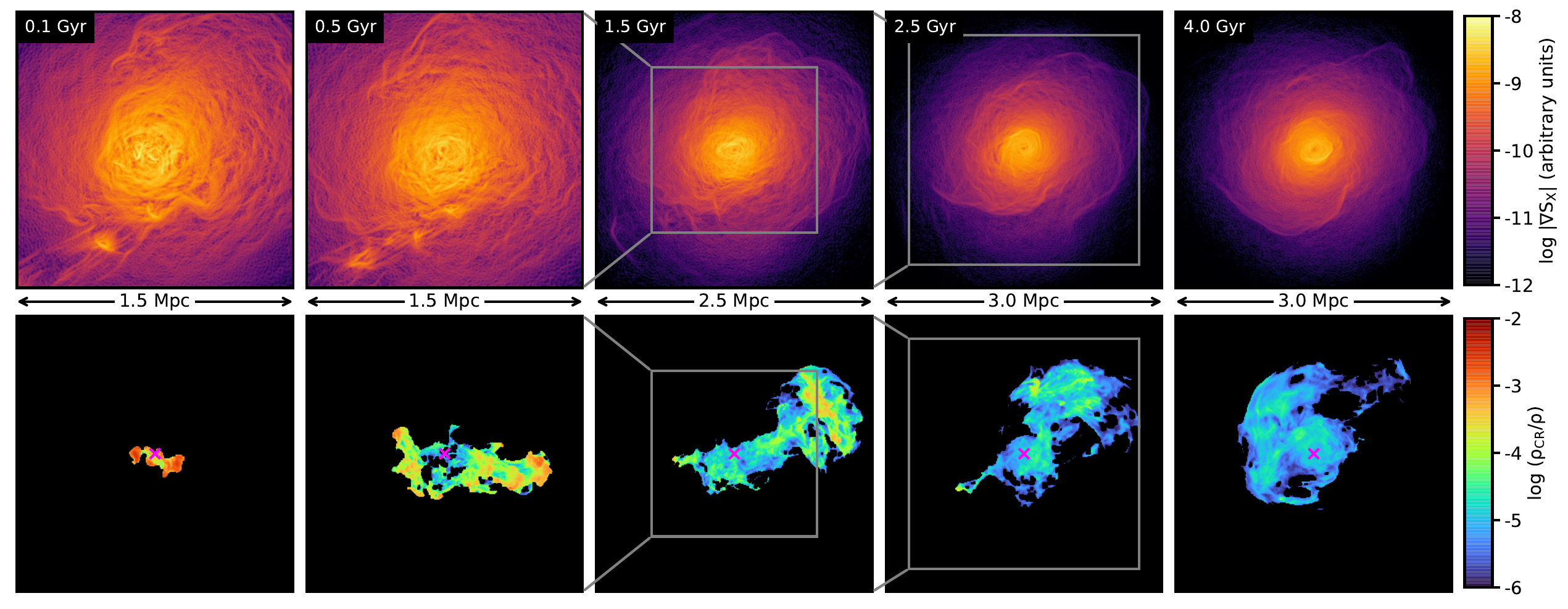}
\caption{The evolution of the ``Merger2''/jet simulation for the first 4~Gyr
after jet ignition, with jets aligned along the $x$-axis. The top panels show
projected X-ray surface brightness gradient, and the bottom panels show projected CR
fraction. The physical width of each panel expands from left to right, indicated
by the scale bars and the inset boxes. The magenta ``x'' marks the cluster
center in the bottom panels. All projections are along
the $z$-axis.\label{fig:R3_b500_jets_x}}
\end{figure*}
    
\begin{figure*}
\centering
\includegraphics[width=0.98\textwidth]{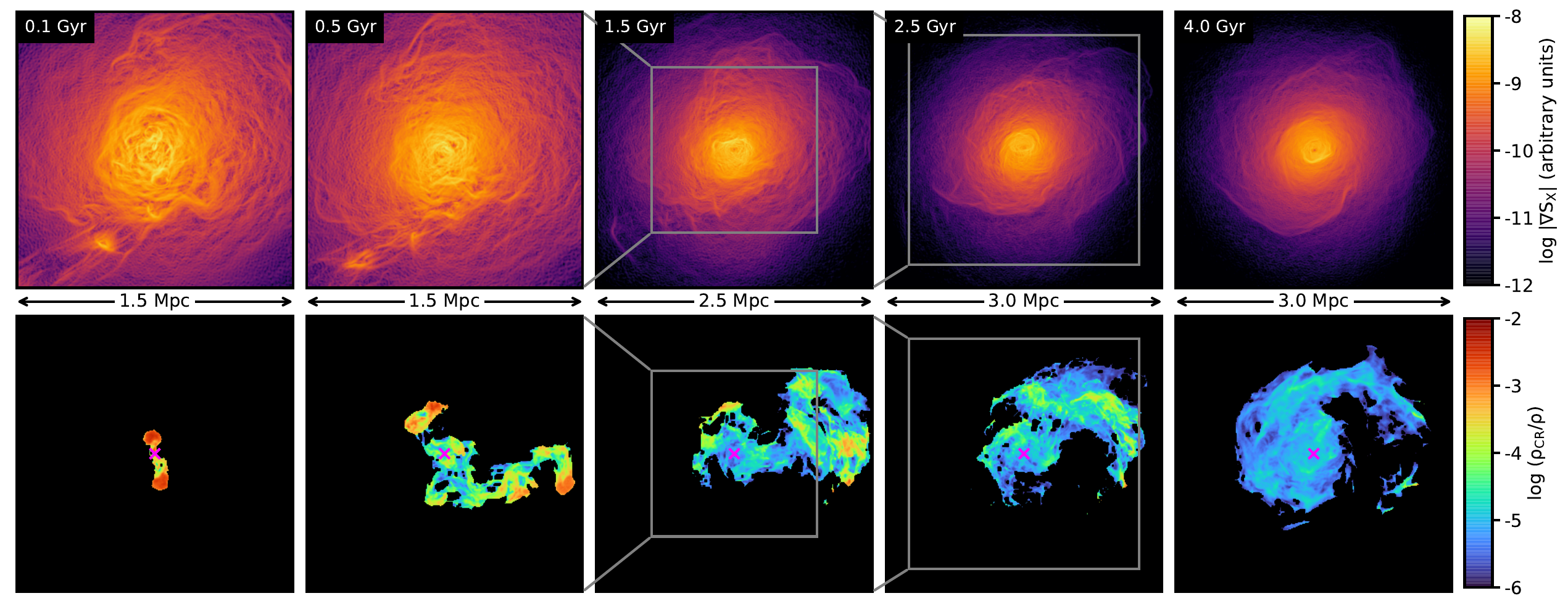}
\caption{The evolution of the ``Merger2''/jet simulation for the first 4~Gyr
after jet ignition, with jets aligned along the $y$-axis. The top panels show
projected X-ray surface brightness gradient, and the bottom panels show projected CR
fraction. The physical width of each panel expands from left to right, indicated
by the scale bars and the inset boxes. The magenta ``x'' marks the cluster
center in the bottom panels. All projections are along
the $z$-axis.\label{fig:R3_b500_jets_y}}
\end{figure*}
  
\begin{figure*}
\centering
\includegraphics[width=0.98\textwidth]{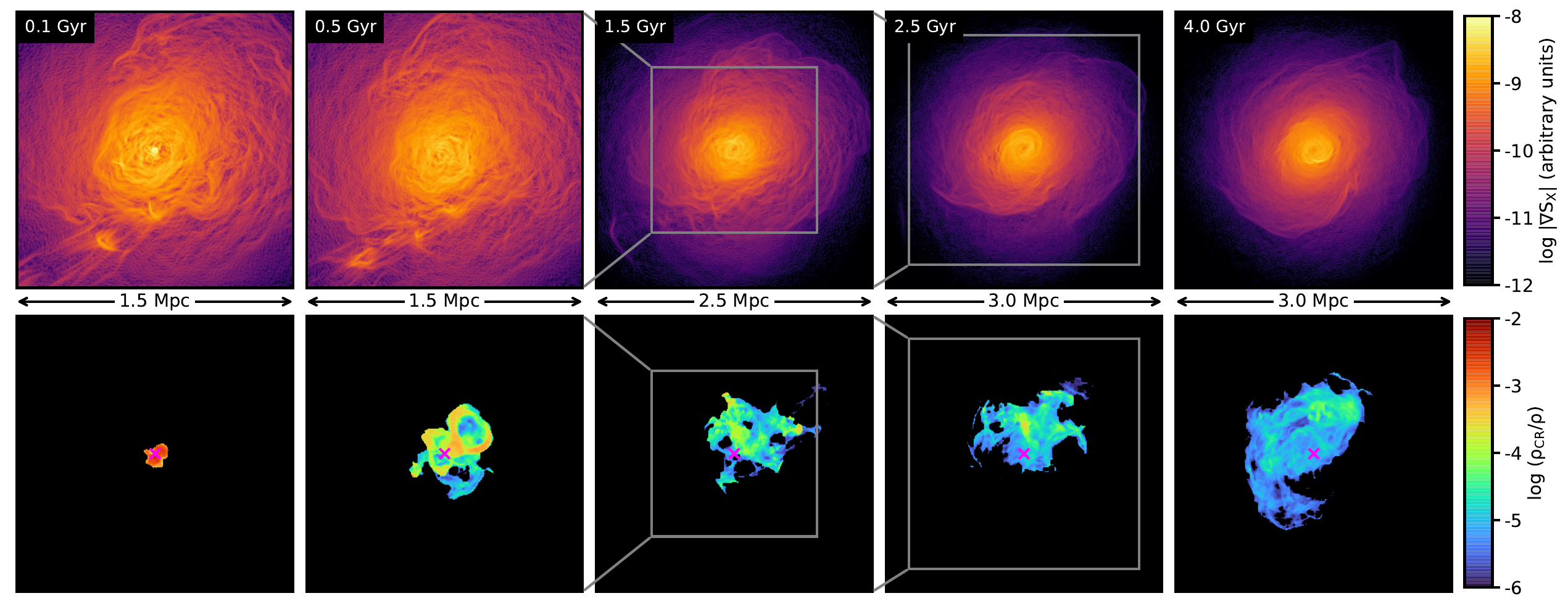}
\caption{The evolution of the ``Merger2''/jet simulation for the first 4~Gyr
after jet ignition, with jets aligned along the $z$-axis. The top panels show
projected X-ray surface brightness gradient, and the bottom panels show projected CR
fraction. The physical width of each panel expands from left to right, indicated
by the scale bars and the inset boxes. The magenta ``x'' marks the cluster
center in the bottom panels. All projections are along
the $z$-axis.\label{fig:R3_b500_jets_z}}
\end{figure*}
  
\begin{figure*}
\begin{center}
\includegraphics[width=0.98\textwidth]{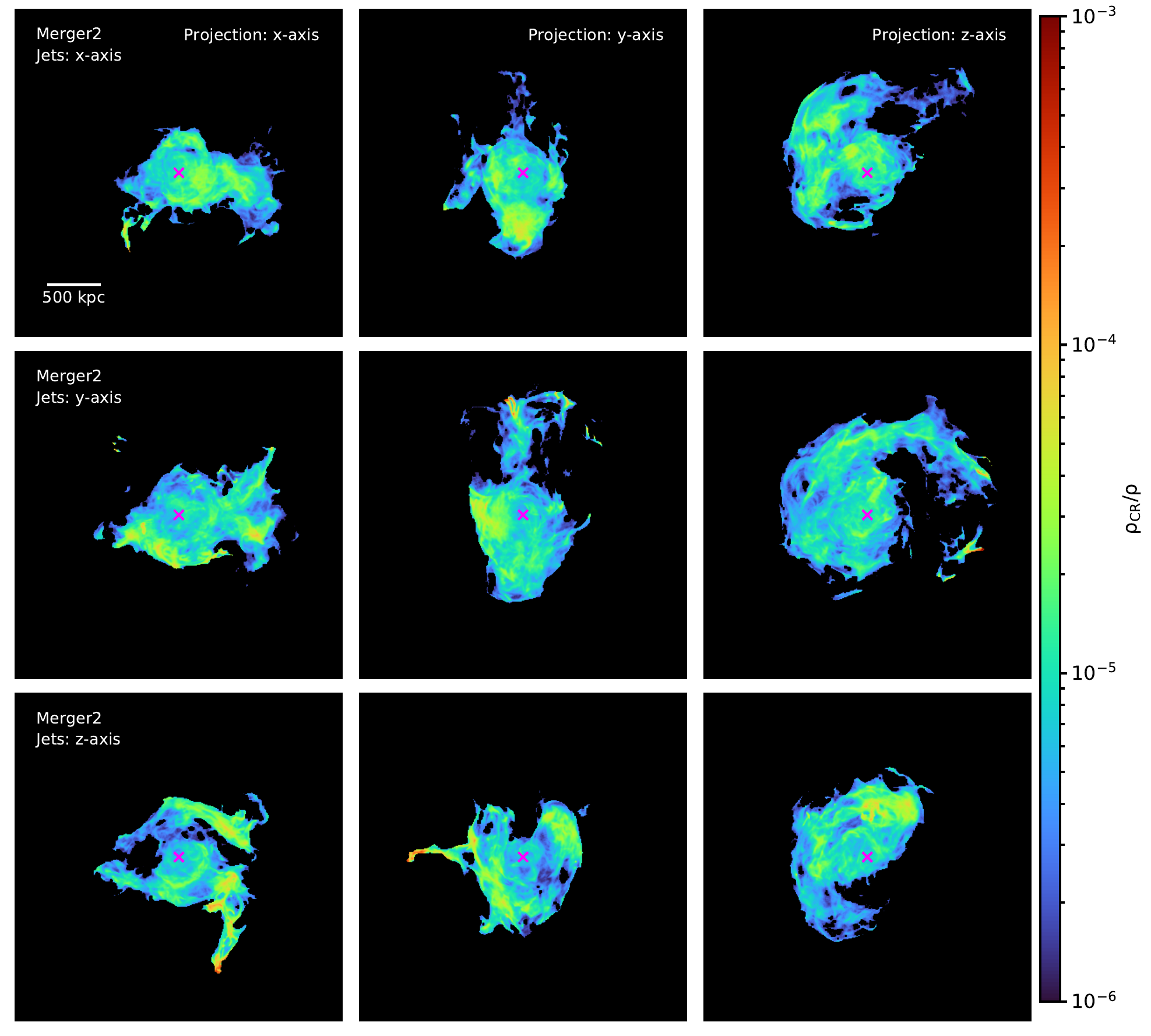}
\caption{The projected CR distribution along the principle axes of the
simulation box for the ``Merger2''/jet simulations, for all cases of the jet
direction, at $\tau$ = 4~Gyr. The position of the cluster center is marked by a
magenta ``x''.\label{fig:R3_b500_cr_proj}}
\end{center}
\end{figure*}

Figure \ref{fig:R5_b500_cr_proj} shows the bubble distribution projected along the
three principal axes of the ``Merger1'' simulations. Along all three projections,
the core region inside $r \sim 200$~kpc is filled nearly uniformly with CRs, with
filamentary structures with sharp boundaries extending out to roughly
$\sim$500-700~kpc in various directions. The width of these extensions clearly
depends on the projection angle. 

Not surprisingly, it is easiest to produce these strip-like features in
the two simulations where the jets are launched within the plane of the merger. 
However, the strips of bubble material are still produced in the simulation where 
the jets are launched perpendicular to the merger plane. This is because the
sloshing motions do not exist solely in the merger plane, but extend perpendicular
to this plane in a ``barrel''-like shape \citep{kes12}. As the jet material travels
outward, it still gets caught up in these motions and is spread around in roughly 
the same fashion. Interestingly, the strips produced in this case are most obvious
in projections observed along either the $x$ or $y$-axes (bottom row, left and center 
panels of Figure \ref{fig:R5_b500_cr_proj}).

\subsubsection{The ``Merger2''/jet Simulations}

The evolution of the jet-injected material in the ``Merger2'' simulations is
shown in Figures \ref{fig:R3_b500_jets_x} (jets in the $x$-direction),
\ref{fig:R3_b500_jets_y} (jets in the $y$-direction), and \ref{fig:R3_b500_jets_z} 
(jets in the $z$-direction). In these simulations, the bubbles are disrupted far 
more quickly than in the ``Merger1'' simulation by the faster and more turbulent 
motions. This leads in the long term to a more diffuse distribution of the bubble 
material than in the ``Merger1'' simulations. There is a very fast cold flow largely 
in the $+x$-direction which begins to the southeast of the core and ends west of it 
before it curves up and eventually counterclockwise around the core. In the
simulations where the jets are lauched within the merger plane (Figures 
\ref{fig:R3_b500_jets_x} and \ref{fig:R3_b500_jets_y}), by $\tau = 0.5$~Gyr the 
bubble material has been caught up within this flow such that it has spread nearly 
$\sim$0.7~Mpc in the western direction, whereas it has barely expanded on the other 
side of the core in the eastern direction. This is even true in the case of the 
simulation with the jets in the $y$-direction, where the bubble launched to the south 
has been essentially immediately pushed 90$^\circ$ from its original direction of 
motion by the fast flow. By $\tau \sim 1.5$~Gyr in both simulations, the bubble
material has spread westward out to a radius of $\sim$1.2~Mpc, and northward by a 
distance of $\sim$0.8~Mpc. In the simulation where the jets are launched 
perpendicular to the merger plane (Figure \ref{fig:R3_b500_jets_z}), the jet material 
is instead spread around in a very chaotic fashion, and does not get caught up into 
this flow until much later.

At later times, from $\tau \sim 2.5-4$~Gyr, the bubble material is completely
caught up in the large-scale bulk motions in the northwest direction from the core,
which spread them out in a large wide arc, with a span of $\sim$1-1.5~Mpc, in a
spiral pattern. These are again confined by the large, sloshing-type motions in
this direction, so they have mostly well-defined edges. However, the
distribution of bubble material is more diffuse and spread out within these
regions than in the ``Merger1'' simulations. These arcs are most 
well-developed in the simulations where the jets are launched within the merger
plane. In the simulation where they are launched in the $z$-direction, the arc-like
spread of bubble material does not appear until very late times ($\tau \sim 4$~Gyr)
and is not as well-developed.

Figure \ref{fig:R3_b500_cr_proj} shows the bubble material distribution
projected along the three principal axes of the ``Merger2'' simulations. All
projections show that this distribution is more diffuse than in the ``Merger1''
simulations. The outward sharp boundaries of the CR distribution are ubiquitous, 
though less pronounced in the case where the jet is launched perpendicular
to the merger plane (bottom row of Figure \ref{fig:R3_b500_cr_proj}).

\subsubsection{The ``Merger1''/bubble Simulation}
  
\begin{figure*}
\centering
\includegraphics[width=0.98\textwidth]{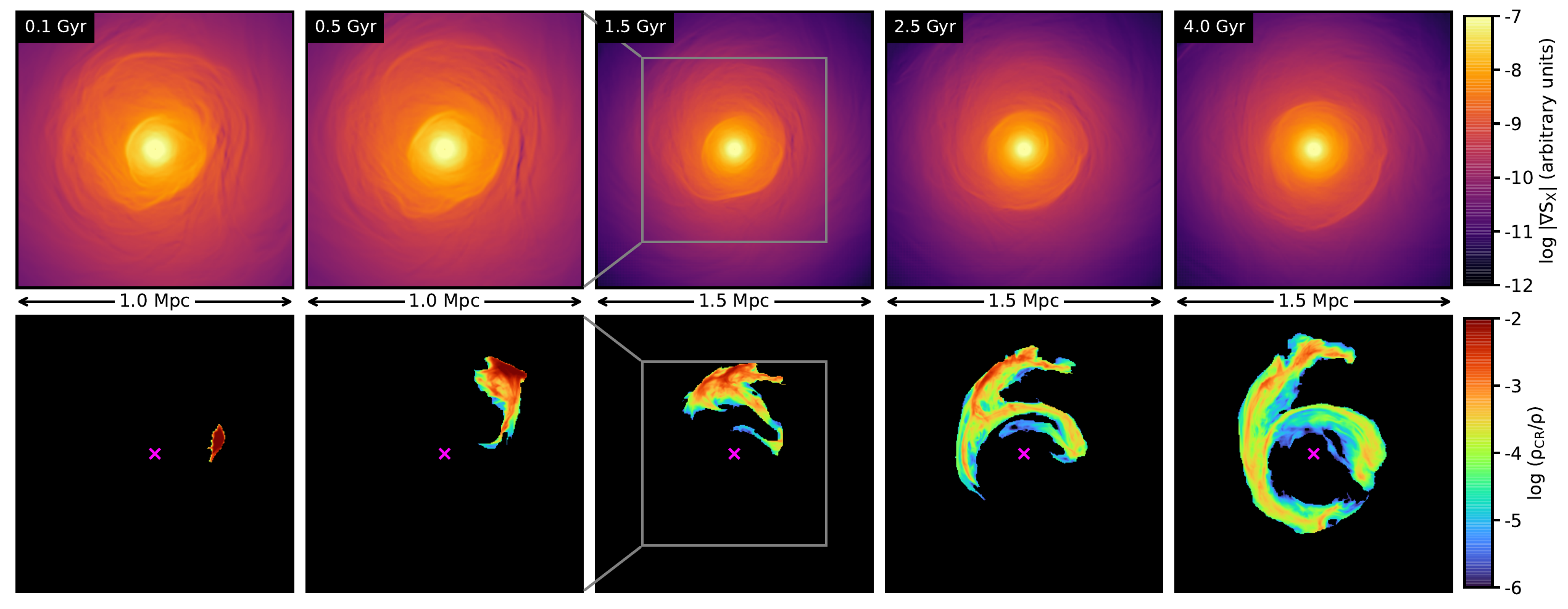}
\caption{The evolution of the ``Merger1''/bubble simulation for the first 4~Gyr
after bubble placement. The top panels show projected X-ray surface brightness gradient, and
the bottom panels show projected CR fraction. The physical width of each panel
expands from left to right, indicated by the scale bars and the inset boxes. The
magenta ``x'' marks the cluster center in the bottom panels. All projections are
along the $z$-axis.\label{fig:200kpc_y}}
\end{figure*}

As mentioned in Section \ref{sec:flash_bubbles}, in this simulation a single
high-entropy bubble is inserted at a radius of 200~kpc from the cluster center, perhaps
injected into the ICM by a local radio galaxy. The evolution of this simulation
is shown in Figure \ref{fig:200kpc_y}. The bubble begins to rise bouyantly and
flattens out in the tangential direction. It becomes quickly caught up in the
sloshing motions as it rises, so that by $\tau \sim 1.5$~Gyr it has already
spread out over a tangential distance of $\sim$~0.75~Mpc. At this epoch, a small
amount of the bubble material, which has become effectively mixed with
lower-entropy gas originally dragged upward by the bubble, begins to fall back
toward its original radius, and spreads around in the tangential direction. The
still-higher entropy bubble material continues to rise, but continues to spread
over a large tangential distance. Lower-entropy material has settled at a radius 
of $r \sim 200-400$~kpc, and has nearly spread around in a complete circle by $\tau
\sim 4$~Gyr.

\begin{figure*}
\begin{center}
\includegraphics[width=0.98\textwidth]{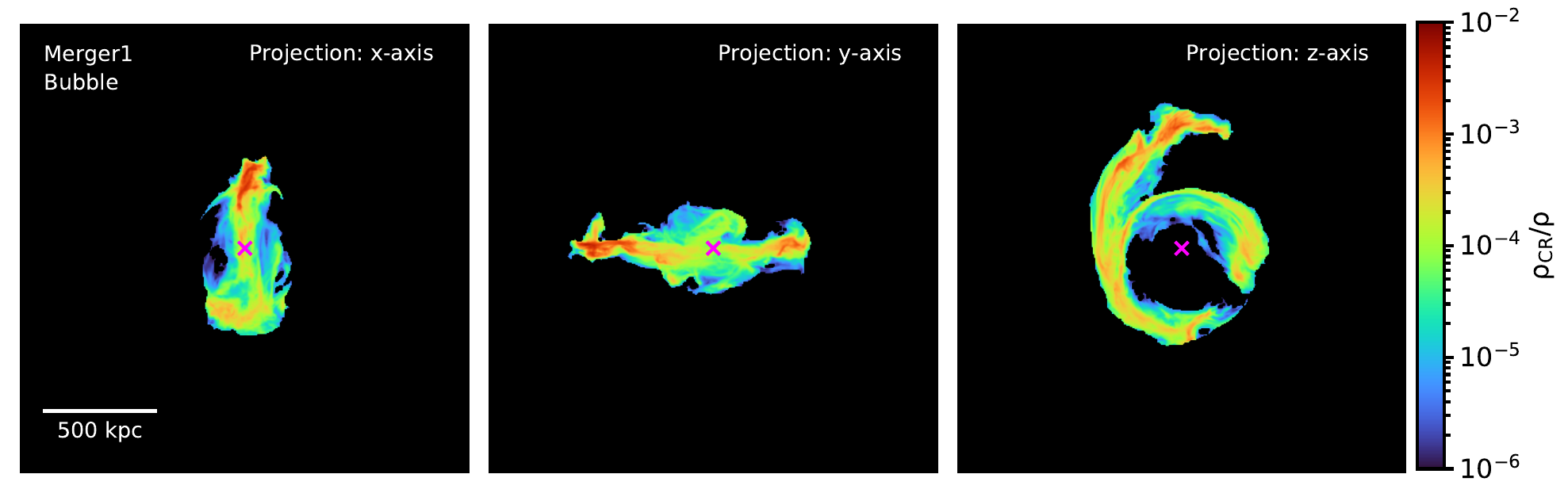}
\caption{The projected CR distribution along the principle axes of the
simulation box for the ``Merger1''/bubble simulation at $\tau$ = 4~Gyr. The
position of the cluster center is marked by a magenta
``x''.\label{fig:200kpc_y_cr_proj}}
\end{center}
\end{figure*}

Figure \ref{fig:200kpc_y_cr_proj} shows the projected bubble material along the
three coordinate axes of the simulation domain at the epoch $\tau = 0.4$~Gyr. Along 
the $x$ and $y$-projections of the simulation, the material dispersed from the 
original bubble appears as a very thin line of width $\sim$100~kpc, nearly $\sim$1~Mpc
in length, except in the core region of radius $\sim$200~kpc where the bubble material 
is starting to be spread around more uniformly. 

In this setup, the bubble originally appears undisrupted at large radius with a high 
concentration of radio plasma. In contrast, in the jet simulatons, the bubbles are 
launched near the cluster center and undergo disruption and mixing with the ICM plasma 
before they reach the same radius. The result is that in the former simulation, the 
density of bubble material which is spread around is higher at large radius than in 
the latter. The two cases differ in morphological detail, but produce common qualitative 
features in the CR distribution --- the ubiquitous elongated, tangential arc-like 
filaments with well-defined sharp boundaries, located much farther in the cluster 
periphery than the initial injection sites. We showed results for only one set of initial
bubble radii for each simulation; for other radii, the picture is similar but with 
different linear scales. For jets oriented at different angles in the plane of the sky 
(along x and y axes), there is no qualitative difference in the final CR distribution.

\subsection{The Distribution of the Magnetic Field in the CR-populated Region}

Radio relics are usually polarized \citep{ens98,vanweeren10,vw19}, with the
polarization direction perpendicular to the long axis of the relic. It is an
open question whether or not such magnetic field alignments arise naturally from
the process of forming the relic itself, or if relics appear predominantely in
regions with magnetic fields oriented tangentially to their surfaces. We can
check the orientation of the magnetic field within the elongated CRe regions in
our simulations. 

In Figure \ref{fig:mag_fields} we show the projected CR distribution along the
$z$-axis of four simulations, with plane-of-sky projected magnetic field
vectors overlaid. The projected magnetic field has been weighed by a factor
$B^2\rho_{\rm CR}$, which will be proportional to the monochromatic emission of
a radio relic with $\alpha = 1$ for $I(\nu) \propto \nu^{-\alpha}$. In each of
these cases, elongated regions of enhanced CR distribution are associated with
magnetic field lines which are largely aligned lengthwise along these features.
Such alignments naturally arise due to the fact that the same motions which
spread the CRs out over long distances stretch and amplify the magnetic field
along the same direction. These magnetic field configurations could be the
origin of polarization in these sources, which would be augmented by the passing of 
weak shock to provide additional field compression and alignment.
      
\section{Summary}\label{sec:summary}

\begin{figure*}
\centering
\hspace{-12mm}
\includegraphics[width=0.43\textwidth]{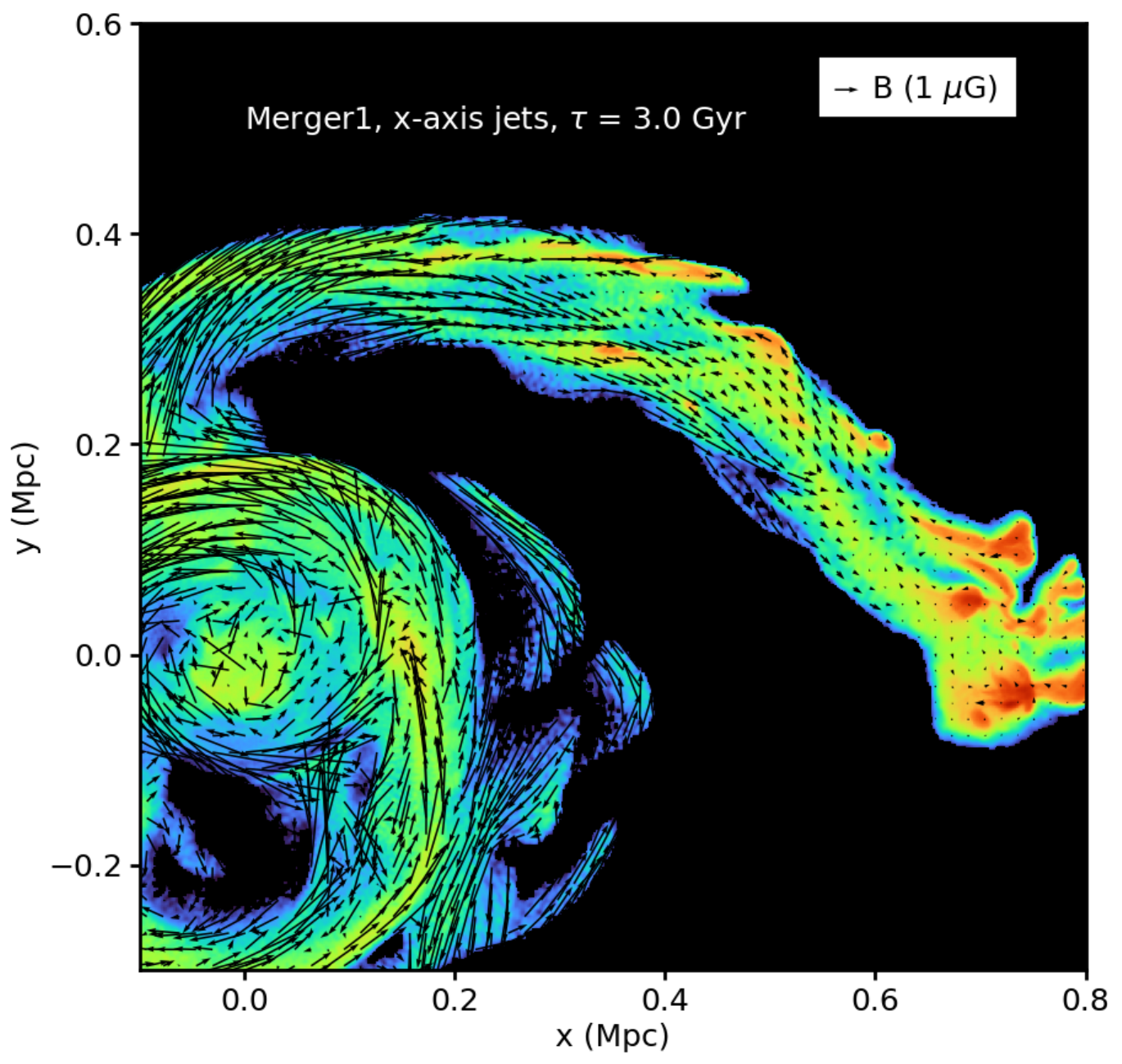}
\includegraphics[width=0.43\textwidth]{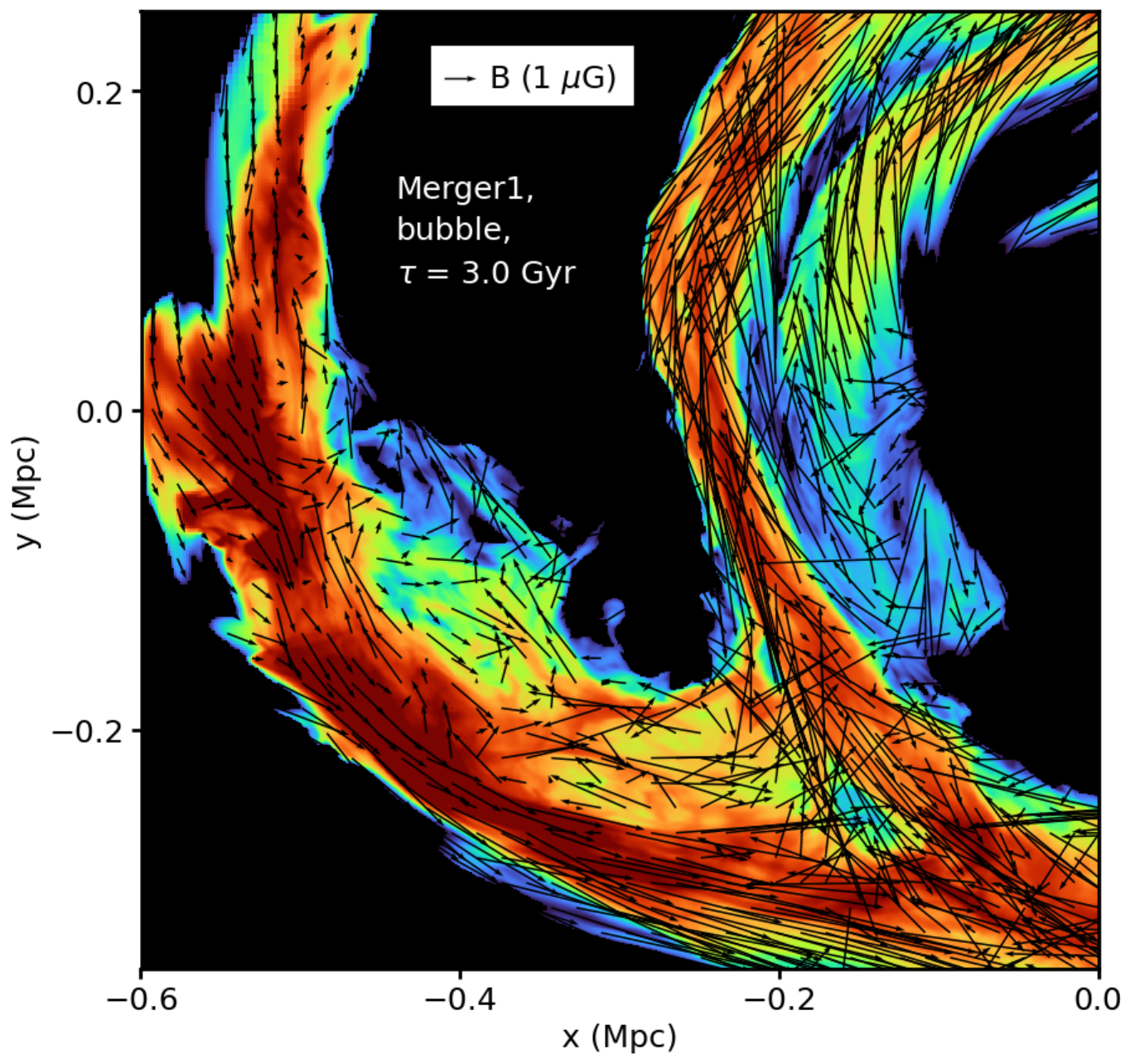}
\includegraphics[width=0.43\textwidth]{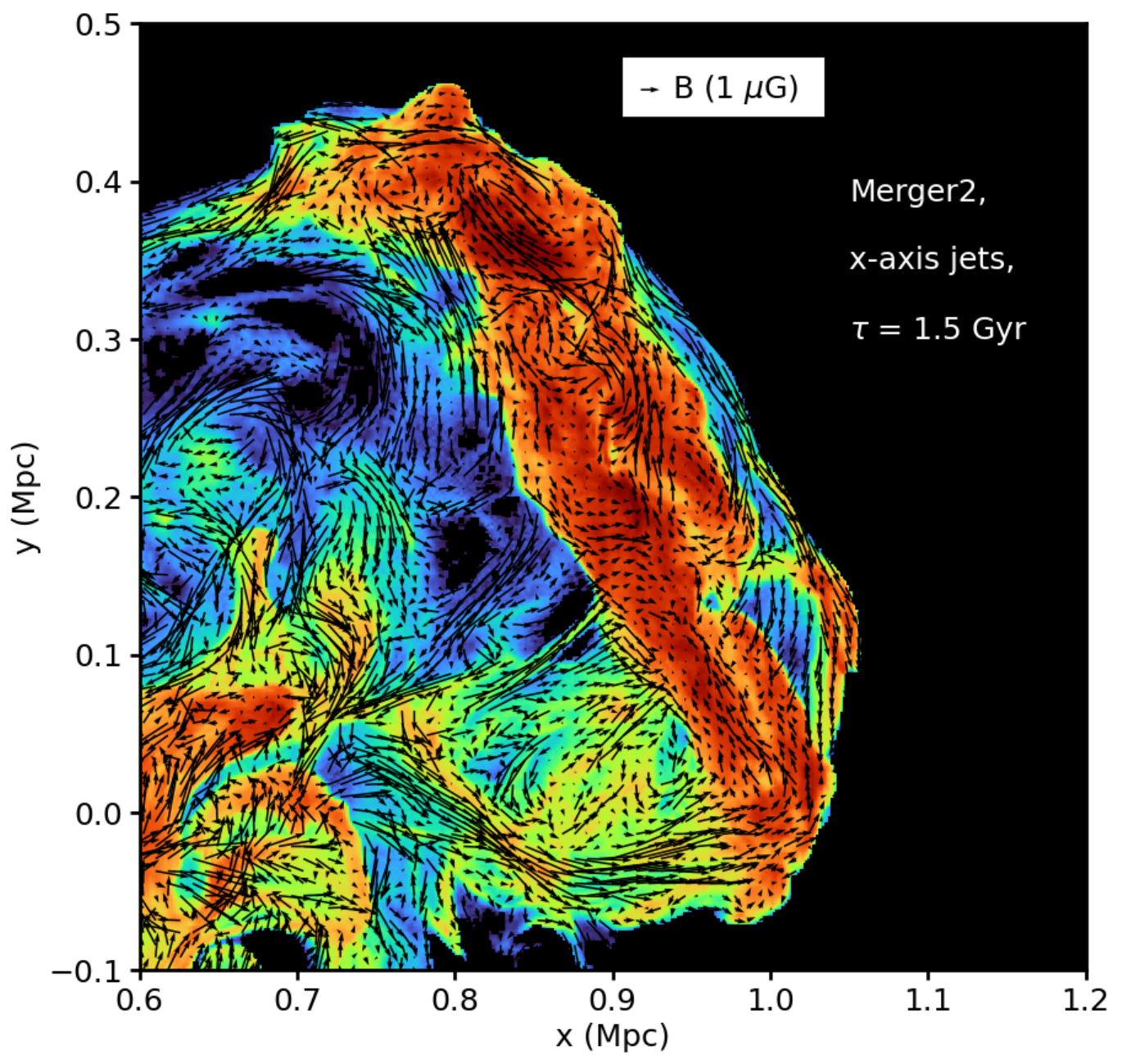}
\includegraphics[width=0.49\textwidth]{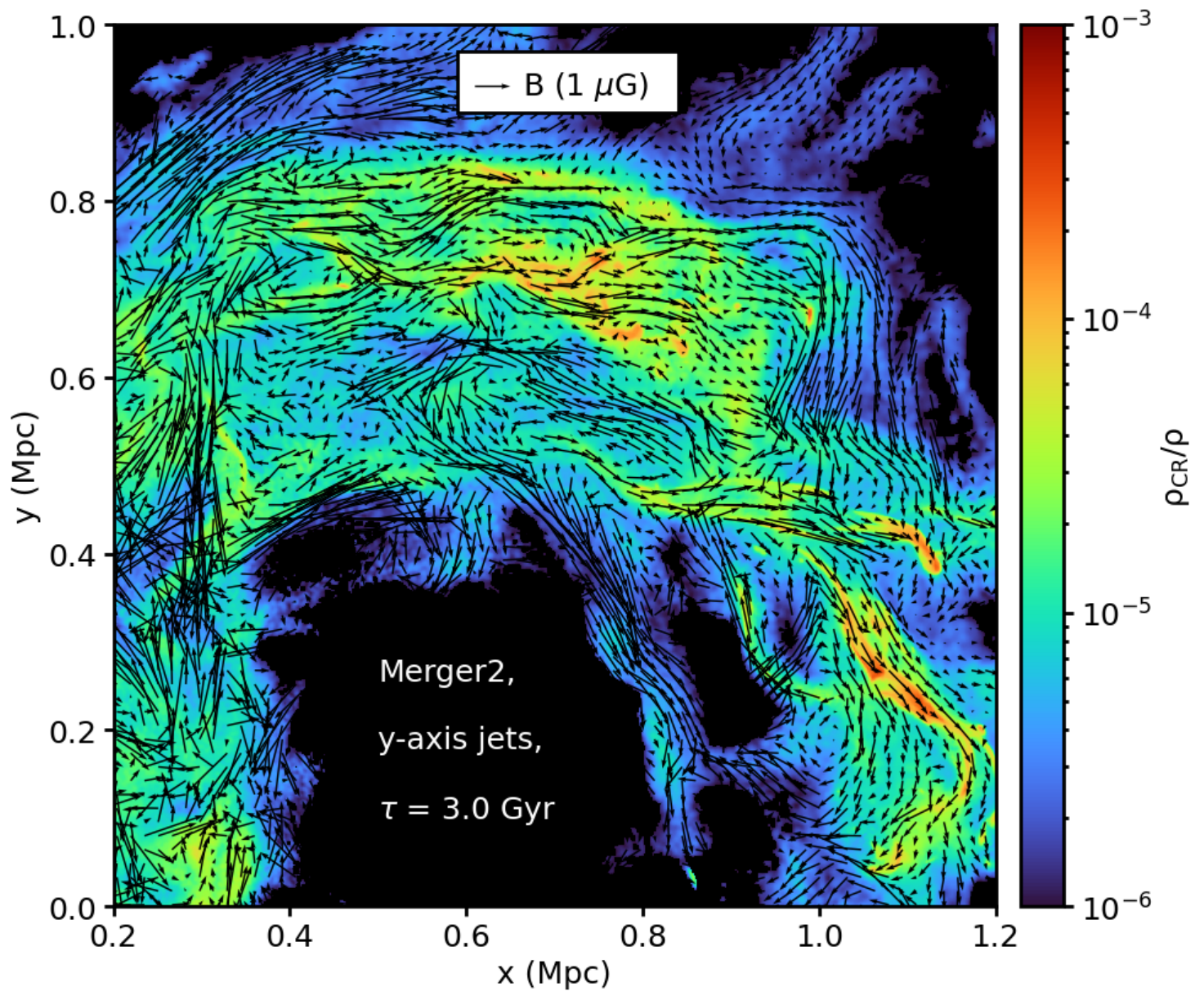}
\caption{Projected CR distribution and plane-of-sky magnetic field vectors in
the vicinity of three extended CR regions in four simulations. The magnetic
field vectors are weighted proportional to $B^2\rho_{\rm CR}$.
\label{fig:mag_fields}}
\end{figure*}

This paper represents an initial exploratory step using MHD simulations of
galaxy cluster mergers with AGN bubbles in testing the hypothesis that the
CRe from AGN-blown bubbles can be turned into radio relics, using a simple
setup of a single AGN outburst creating bubbles in already well-developed merger-driven
gas motions. Specifically, this occurs by the transport of the CRe by merger-driven 
gas motions to larger radii where they are stretched in the direction tangential 
to the cluster equipotential surfaces / isophotes. Such CRe can then by reaccelerated 
by a passing merger shock, explaining the spatial association of the radio features
with X-ray detected shocks in some clusters but also explaining why not all
radio relics have such associations. In particular, we find that:

\begin{itemize}

\item The bubbles produced by jets which emanate from a central AGN are quickly 
offset (in our simulations, within 0.5~Gyr) from the original jet axis by sloshing 
gas motions. Faster and more turbulent motions from a more violent merger can offset 
bubbles to a greater degree, but these motions also tend to disrupt bubbles rather 
quickly. 

\item Cluster mergers which produce relatively gentle sloshing motions can
spread bubble material out to relatively large radii, of roughly $r \sim
0.5-0.6$~Mpc. More energetic mergers, which can produce fast-moving bulk
motions which quickly develop into turbulence, can spread it to even larger
radii, of $\sim 1$~Mpc. 

\item The CRe structures produced by these motions in either case can be stretched
greatly in the tangential direction, $\sim$1~Mpc, especially if the merger
occurs with a significant impact parameter, such mergers produce fast
motions in the tangential direction. Such spatial distribution is expected in a 
stratified, largely centrally-symmetric cluster atmosphere (which is true even for 
highly disturbed clusters) --- a parcel of gas with a given specific entropy spreads 
much more easily along the isentropic surface than in the radial direction. These 
same motions will stretch and amplify magnetic fields parallel to these structures, 
which could potentially explain the polarization of radio relics. 

\item In mergers where the gas motions are comparatively gentle and less
turbulent, such as our ``Merger1'' simulation, the long tangential regions of
enhanced bubble material are very narrow, similar to those seen in many radio relics.
In the more turbulent ``Merger2'' simulation, the regions of enhanced CRe are
unsurprisingly wider and more diffuse. Of course, determining whether or not the full 
region populated by CRe is lit up after a shock passage will require more detailed
simulation, including the energetics of the CRe (see below), so it is premature
to suggest that sloshing motions produce thin relics and more turbulent motions produce 
thicker ones. 

\item The production of the long, tangential features is relatively insensitive to
the jet direction, even when it is launched perpendicular to the plane of the merger which
initiated the gas motions, since these motions extend also in this perpendicular direction. 
However, in this extreme case of a jet launched perpendicular to this plane, the extended 
features may take longer to develop and/or depend on projection angle. If the gas motions
are very turbulent, then in this particular case the extended features may not appear.

\item In our simulation which used a single bubble at a radius of 200~kpc, the density of 
bubble material is much larger and it is distributed in more coherent structures than in 
the simulations where the bubbles are produced by an AGN at the cluster center. This may 
indicate that thinner and more coherent radio features may be produced by radio galaxies 
at larger radii which inject CRe locally, but future simulation works with more detailed 
physics are required to examine this possibility. These simulations should also include 
more realistic inflation of bubbles, since real bubbles will be born with kinetic energy, 
momentum, and an internal magnetic field structure, which the method employed in the 
\code{FLASH} simulation does not include. 

\item The core region of the cluster in the jet simulations is quickly filled up with CRe 
which originated from the AGN, confirming the hypothesis that the central AGN
could provide the seed electrons for radio mini-halos \citep[see e.g.][]{zuh13}.

\end{itemize}

Such a hypothesis for the origin of radio relics neatly explains several
puzzling features:

\begin{enumerate}
  
\item they are not always seen coincident with merger shocks on the sky, since
the shock may have run over the CRe-populated region and left, though it should
still be nearby given the short cooling times of relativistic electrons

\item why merger shocks do not always produce detectable radio relics, since it is 
required that the shocks will pass over CRe-populated regions

\item why relics do not always have shapes that resemble merger shocks; in this
case, the shape of the relic is simply the shape of the CRe-populated region,
which depends on how they were distributed by gas motions. 

\end{enumerate}

Needless to say, this is a very simple study which only scratches the 
surface of the possible scenarios in which radio plasma originatig in AGN bubbles 
can be spread around by gas motions. A much larger parameter space needs to be 
explored in the following areas:

\begin{itemize}

\item Observationally, the bubble energy of $10^{61}$~erg (created in our jet
simulations) corresponds to the most energetic bubbles observed in clusters, 
which may not be typical of the history of many clusters with radio relics. 
Radio jets with less power will generate bubbles with less energy and 
smaller size, which will be less capable of bouyantly rising to large 
radii in the ICM, and their material must be transported there by gas 
motions instead, as hypothesized here. This shows why the second scenario we 
explored, that of radio plasma injected at larger radius, is highly relevant
and deserves further study. 

One of the features of bubble simulations such as ours is that as they 
rise, they are shredded by Kelvin-Helmholtz and Rayleigh-Taylor instabilities,
and this shredding can occur more quickly if the medium is turbulent. Instead
of disrupting a bubble uniformly all at once, what often happens is that the 
bubble separates into ``pockets'' of coherent radio plasma, some of which may 
last longer than others (for example, the large coherent structure in the 
bottom-left panel of Figure \ref{fig:mag_fields} originated from a pocket of
radio plasma which split off in a very turbulent medium, as seen in the first
few epochs in Figure \ref{fig:R3_b500_jets_x}). In a more turbulent medium, 
smaller bubbles will get disrupted into smaller pockets, or may be more easily
mixed into the ICM very quickly before rising very far, in which case one would
not see the coherent structures at large radii shown in these simulations.

Aside from these caveats, it is important to note that since this study assumes that 
the radio plasma from the bubble has become mixed within the ICM, the features which 
produce radio relics are not fresh CRe but rather those that have been present 
for some time and have cooled to lower energies where they are not radiating.
Re-acceleration of these CRe by a shock to radio-emitting energies is still 
essential for radio relics to be observed in our scenario. Our study is thus  
distinct from the situation where direct acceleration of plasma in radio lobes 
occurs \textit{before} it has mixed in with the ICM, as simulated recently by 
\citet{nolting2019a,nolting2019b}. The present work only considers how gas motions 
could influence the spatial distribution of CRe to provide the seeds for later 
re-acceleration.

\item The model employed in this work for tracing the CRe was very simple--a 
tracer which was advected along with the fluid, initially set to unity for the 
material ejected by the jet or within the initial bubble. A more accurate model 
would treat the physics of the CRe explicitly, allowing for (re)acceleration by 
shocks and/or turbulence, cooling from radiative losses and Coulomb collisions, 
and/or diffusion and streaming \citep[as in][]{zuh13,don14,zuh15,yang17,winner19,ogro20,dom20}. 
This would provide a way to more definitively determine if the energetic properties 
of the CRe can produce radio emission with a brightness and spectrum consistent 
with what is seen in observations, depending on the shock strength and local 
magnetic field. 

\item It is also true that we have limited ourselves to a single AGN outburst
in these simualations, as well as a single merger event producing gas motions. In 
reality, minor mergers will be producing gas motions semi-frequently, and (more 
importantly) several AGN duty cycles are likely to occur within the timescale for 
the development of gas motions, complicating the simple picture we have presented here. 
Whether including multiple AGN outbursts and/or multiple mergers will produce 
multiple arcs of CRe which could be re-accelerated later to produce radio relics, or if the 
CRe would simply be mixed together in a more irregular fashion, remains to be seen and 
is probably dependent on the exact scenario. Additionally, we only examined a single 
case where the AGN outbursts occurred only after the gas motions were well-developed. 
A situation where the bubbles are produced and begin to rise first and a merger occurs
afterward should also be explored, though because our hypothesis relies on the mixing 
of the bubble material with the ICM and is then subsequently transported by gas motions,
we might expect our results to hold.

\end{itemize}

From these considerations, obvious follow-ups to this work suggest themselves. 
First, idealized simulations such as those presented here which vary the timing
and frequency of AGN outbursts and mergers should be explored. Second, more complex
physics for the evolution of the CRe should also be included. Finally, we should 
follow-up a simulation such as those shown here by merging the cluster with CRe arcs
with another cluster, allowing the CRe to be passed over by a merger shock. Perhaps the 
most promising path forward for a full investigation of this hypothesis in the most 
general terms would be to include the effects of AGN feedback, acceleration by shocks 
and turbulence, and radiative losses within a cosmological simulation (where the 
merger-driven gas motions come ``for free''), as was done recently by \citet{vazza2021},
so that a large sample of possible outcomes could be studied, potentially producing relics
with different sizes, shapes, locations, and brightnesses. This would provide the best 
way to move from the provisional hypothesis put forward in this work, namely that it is 
\textit{possible} to produce radio relics using CRe produced by the central AGN, to 
the question of under what circumstances this actually occurs.
    
\acknowledgments

We thank the anonymous referee for their comments and suggestions, which have 
improved this manuscript. JAZ and PN acknowledge support from the Chandra X-ray Center, 
which is operated by the Smithsonian Astrophysical Observatory for and on behalf of NASA 
under contract NAS8-03060. 


\facilities{Pleiades (NASA/Ames Research Center), Hydra (Smithsonian Institution)}

\software{AREPO\footnote{\url{https://arepo-code.org}} \citep{springel10,weinberger2017}
          AstroPy\footnote{\url{https://www.astropy.org}} \citep{ast13},
          FLASH\footnote{\url{http://flash.uchicago.edu}} \citep{fry00,dub09,lee09,fry10},
          Matplotlib\footnote{\url{https://matplotlib.org}} \citep{hun07},
          NumPy\footnote{\url{https://www.numpy.org}} \citep{vdw11,harris20},
          yt\footnote{\url{https://yt-project.org}} \citep{tur11}}


\bibliographystyle{aasjournal}

\end{document}